\documentclass[journal]{vgtc}                
\ifpdf
  \pdfoutput=1\relax                   
  \usepackage{graphicx}                
  \DeclareGraphicsExtensions{.pdf,.png,.jpg,.jpeg} 
\else
  \usepackage{graphicx}                
  \DeclareGraphicsExtensions{.eps}     
\fi%

\graphicspath{{figures/}{pictures/}{images/}{./}} 

\usepackage{microtype}                 
\PassOptionsToPackage{warn}{textcomp}  
\usepackage{textcomp}                  
\usepackage{mathptmx}                  
\usepackage{times}                     
\usepackage{cite}                      
\usepackage{tabu}                      
\usepackage{subcaption}
\usepackage{booktabs}                  
\usepackage[dvipsnames]{xcolor}

\usepackage{algorithm}
\usepackage{algpseudocode}

\definecolor{requirement-light}{rgb}{0.98,0.98,0.98}
\definecolor{requirement-dark}{rgb}{0.98,0.73,0.01}
\newenvironment{requirement-leftbar}[1][\hsize]
{%
	\MakeFramed{\hsize#1\advance\hsize-\width\FrameRestore}%
}

\definecolor{cb_orange}{rgb}{1.0,0.51,0.0}
\definecolor{cb_blue}{rgb}{0.22,0.49,0.72}
\definecolor{cb_green}{rgb}{0.3,0.67,0.29}
\definecolor{cb_red}{rgb}{0.89,0.1,0.11}
\definecolor{cb_purple}{rgb}{0.6, 0.31, 0.64}
\definecolor{cb_brown}{rgb}{0.6, 0.4, 0.2}
\definecolor{cb_crimson}{rgb}{0.86, 0.08, 0.24}

\preprinttext{Submitted to IEEE VIS: Visualization \& Visual Analytics.}

\onlineid{0}

\vgtccategory{Research}
\vgtcpapertype{Area 2: Applications}

\title{Scope2Screen: Focus+Context Techniques for Pathology Tumor Assessment in Multivariate Image Data}

\author{Jared Jessup$^{1}$, Robert Krueger$^{1}$, Simon Warchol,  John Hoffer, Jeremy Muhlich, Cecily C. Ritch, Giorgio Gaglia,\\ Shannon Coy, Yu-An Chen, Jia-Ren Lin, Sandro Santagata, Peter K. Sorger, Hanspeter Pfister}

\authorfooter{
\item
 $^{1}$: Robert Krueger and Jared Jessup contributed equally to this work.
\item
 Robert Krueger, Jared Jessup, Simon Warchol and Hanspeter Pfister are with the School of Engineering and Applied Sciences, Harvard University.
 Contact: krueger@g.harvard.edu
\item
 Giorgio Gaglia, Cecily C. Ritch, Shannon Coy, and Sandro Santagata are with Brigham and Women's Hospital, Harvard Medical School.
\item
 Robert Krueger, John Hoffer, Jeremy Muhlich, Jia-Ren Lin, Yu-An Chen, Sandro Santagata, and Peter K Sorger are with the Laboratory of Systems Pharmacology, Harvard Medical School. 
}

\shortauthortitle{Biv \MakeLowercase{\textit{et al.}}: Global Illumination for Fun and Profit}

\abstract{Inspection of tissues using a light microscope is the primary method of diagnosing many diseases, notably cancer.  Highly multiplexed tissue imaging builds on this foundation, enabling the collection of up to 60 channels of molecular information plus cell and tissue morphology using antibody staining. This provides unique insight into disease biology and promises to help with the design of patient-specific therapies.  However, a substantial gap remains with respect to visualizing the resulting multivariate image data and effectively supporting pathology workflows in digital environments on screen. We, therefore, developed Scope2Screen, a scalable software system for focus+context exploration and annotation of whole-slide, high-plex, tissue images. Our approach scales to analyzing 100GB images of $10^9$ or more pixels per channel, containing millions of individual cells.  A multidisciplinary team of visualization experts, microscopists, and pathologists identified key image exploration and annotation tasks involving finding, magnifying, quantifying, and organizing regions of interest (ROIs) in an intuitive and cohesive manner. Building on a scope-to-screen metaphor, we present interactive lensing techniques that operate at single-cell and tissue levels. Lenses are equipped with task-specific functionality and descriptive statistics, making it possible to analyze image features, cell types, and spatial arrangements (neighborhoods) across image channels and scales. A fast sliding-window search guides users to regions similar to those under the lens; these regions can be analyzed and considered either separately or as part of a larger image collection.  A novel snapshot method enables linked lens configurations and image statistics to be saved, restored, and shared with these regions. We validate our designs with domain experts and apply Scope2Screen in two case studies involving lung and colorectal cancers to discover cancer-relevant image features.}


\CCScatlist{ 
 \CCScat{K.6.1}{Management of Computing and Information Systems}%
{Project and People Management}{Life Cycle};
 \CCScat{K.7.m}{The Computing Profession}{Miscellaneous}{Ethics}
}

\teaser{
  \centering
  \includegraphics[width=\linewidth]{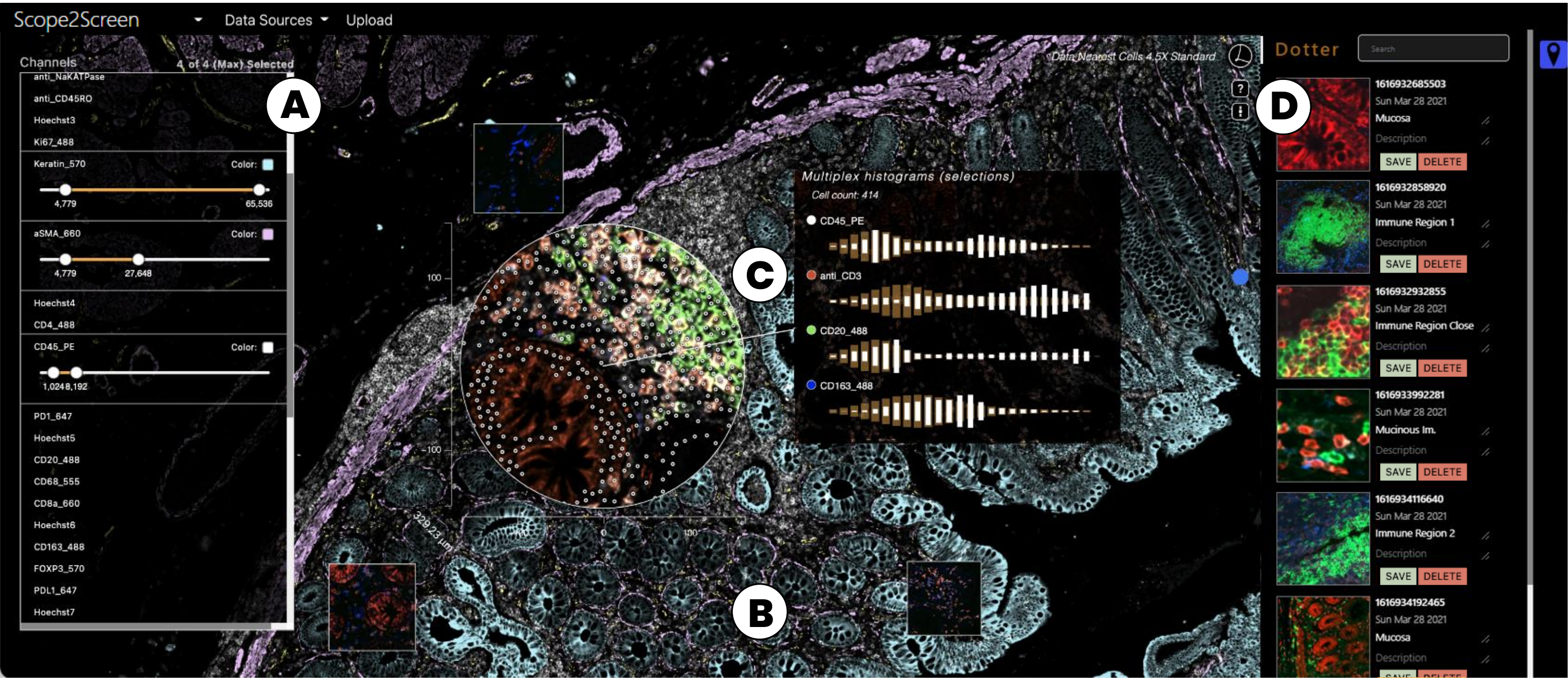}
    \vspace{-1.7em}
  \caption{Scope2Screen offers (A) Channel \& color selection for multi-channel rendering. (B) A WebGL-based viewer capable of rendering 100+GB sized high-plexed and high-resolution ($\ge$ 30k $\times$ 30k) image data in real-time. (C) Interactive lensing for close-up analysis - the lens shows a multi-channel immune setting that is different from the global context highlighting basic tissue composition. (D) Dotter panel - stores and organizes snapshots of annotated ROIs to filter, restore, navigate to the image location.}
  \label{fig:teaser}
}

\begin{document}




\firstsection{Introduction}

\maketitle



Since the end of the 19th century, the diagnosis of many diseases - cancer in particular - has involved human inspection of stained tissue sections using a simple light microscope~\cite{chapman_scope_2020}. Histopathology in both research and clinical settings still involves microscopy-based inspection of physical slides but a rapid shift to digital instruments and computational analysis (\textit{scope to screen}) is now underway~\cite{chapman_scope_2020}. Digital pathology~\cite{weinstein_prospects_1986} in a clinical setting focuses on the analysis of tissues stained with colorimetric dyes (primarily hematoxylin and eosin, H\&E~\cite{titford_long_2005}) supplemented by single-color immunohistochemistry methods that use antibodies to detect molecular features of interest~\cite{pallua_future_2020}. In research settings, recently developed high-plex imaging methods such as  CyCIF~\cite{lin_cyclic_2016, lin_highly_nodate}, CODEX~\cite{goltsev_deep_2018}, and mxIF~\cite{gerdes_highly_2013}  to measure the levels and sub-cellular localization of 20-60 proteins, providing single-cell information on cell identities and states in a preserved tissue environment. The resulting data are complex, involving multi-channel gigapixel images having $10^6$ or more cells. Underdevelopment of analytical and visualization methods is a barrier to progress in digital pathology, explaining the continuing dominance of physical slides.  

Machine learning on high-plex tissue images has shown promise, particularly with respect to automated classification of cell types~\cite{liu_comparison_2019,campanella_clinical-grade_2019}, tissue morphologies~\cite{stoltzfus_cytomap_2020}, and cellular neighborhoods~\cite{madabhushi_image_2016}.  However, such data-driven approaches do not leverage hard-won information known primarily to anatomic pathologists on which cell and tissue morphologies are significantly associated with disease outcome or response to therapy. A hundred years of clinical pathology has investigated many striking and recurrent image features whose significance still remains unknown. A critical need, therefore, exists for new software tools that optimally leverage human-machine collaboration in ways that are not supported by existing  interfaces~\cite{nicholas_sofroniew_naparinapari_2021, allan_omero_2012}.


Pathologists are very efficient at extracting actionable information from physical slides, frequently panning across a specimen while switching between low and high magnifications. They record key observations in notes and by placing dots on slides next to the key features. Digital software needs to reproduce this efficiency and functionality (including a ‘dotting’ function) while using visual metaphors to present associated data and using machine learning to find similar and dissimilar visual fields.  Designing scalable visual interfaces that will work in the context of high-volume clinical workflows~\cite{molin_diagnostic_2016} and to high-dimensional research data represents a substantial challenge.

We addressed these challenges as a team of visualization researchers, pathologists, and cell biologists via a process of goal specification,  iterative testing and design, and real-world implementation in a biomedical research laboratory. We make three primary contributions.  \textbf{(1)} We demonstrate task-tailored, lens-centric focus+context technique, which enables intuitive interaction with large (ca. 100 GB) multi-channel images and linked multivariate data (Fig.~\ref{fig:teaser}). The lensing technique allows users to focus on different aspects of a region for close-up analysis while maintaining the surrounding context. We design novel domain-specific encodings in which features computed from the image (spatial cross-correlation or cell identity) can be accessed in conjunction with the image. \textbf{(2)} We integrate interactive real-time spatial histogram similarity search algorithms able to identify recurrent patterns across gigapixel multi-channel images at different resolutions. Integrated into the lens, this search guides analysts to regions similar to the one in focus, enabling exploratory analysis at scale. \textbf{(3)} We present a scalable system that combines lens and search features with interactive annotation tools, enabling a smooth transition from exploration to knowledge externalization. Analysts can save, filter, and restore regions of interest (ROIs) within the image space (along with underlying statistics of the filtered single-cell data, channel identities, and color settings) and export them for continued study. Two use-cases demonstrate the applicability of our approach to patient-facing (translational) cancer research and point to future applications in diagnosis and patient care.

\section{Related Work}
The related work is three-fold.
We first discuss large-scale image viewers as an enabler for our approach. We then summarize focus+context techniques in comparison to overview+detail and pan\&zoom, with a focus on image data. Lastly, we compare ROI annotation approaches.

\subsection{Scalable Image Viewers For Digital Pathology}
Many biomedical visualization systems focus on the display of large 2D imaging data and apply multi-resolution techniques such as image pyramids~\cite{in_e_1984} to handle large data sizes at interactive rates. DeepZoom~\cite{microsoft_silverlight_2021} hierarchically divides images into tile pyramids and delivers pieces as required by the viewer. Zarr~\cite{alistair_miles_zarr-developerszarr-python_2020, noauthor_zarr_nodate}, a file format and library, abstracts this concept by providing storage of chunked, compressed, N-dimensional arrays. Viewers such as OpenSeadragon~\cite{noauthor_openseadragon_nodate} and Viv~\cite{manz_viv_2020} leverage these libraries and add GPU-accelerated rendering capabilities. On top of that, many solutions offer data-management, atlas, and analysis capabilities. OMERO PathViewer~\cite{allan_omero_2012} is a widely used web-based viewer for multiplexed image data. As an extension to the data management platform OMERO, it supports a variety of microscope file formats.
Online cancer atlases such as Pancreatlas~\cite{saunders_pancreatlas_2020} and Pan-Cancer~\cite{weinstein_cancer_2013} support data exploration with storytelling capabilities. Minerva Story~\cite{hoffer_minerva_2020, rashid_interpretative_2020} is a new tool used to create atlases for the Human Tumor Atlas Network~\cite{rozenblatt-rosen_human_2020}.
Other solutions focus on combining image visualization with analytics. Napari~\cite{nicholas_sofroniew_naparinapari_2021} is a fast and light-weight multi-dimension viewer designed for browsing, annotating, and analyzing large multi-dimensional images. Written in Python, it can be extended with analytic functionality, e.g., in combination with SciMap~\cite{nirmal_et_al_scimap-_nodate}. Other analytical tools focus on end-users, such as the open-source solutions histoCAT~\cite{schapiro_histocat_2017} and Facetto~\cite{krueger_facetto_2020}, and commercial tools such as Halo~\cite{indica_labs_halo_nodate} and Visiopharm's TissueAlign~\cite{tissue_align} supporting split-screen comparison for serial sections. Screenit~\cite{dinkla_screenit_2017-1} presents a design to analyze smaller histology images at multiple hierarchy levels. Similarly, ParaGlyer~\cite{morth_paraglyder_2020} is an analysis approach for multiparametric medical images that permits analysis of associated feature values and comparisons of volumetric ROIs by voxel subtraction.
Somarakis et al.~\cite{somarakis_visual_2021} offer comparison views with a focus on spatially-resolved omics data in a standard viewer. These tools feature multiple linked views for overview+detail exploration. In comparison, our solution focuses on interactive focus+context and rich annotation with contextual details displayed near the ROI and supports a neighborhood-aware similarity search on top of local image pixel and feature value comparison.
Most viewers operate on much smaller datasets. Our viewer builds on Facetto~\cite{krueger_facetto_2020} and Minerva~\cite{hoffer_minerva_2020, rashid_interpretative_2020} and supports multi-channel and cell-based rendering with linked data at a scale few other solutions support. 
The main contribution of this paper, however, is the embedded interactive lensing technique for multivariate image data and its task-tailored features supporting the digital pathology workflow. 

\vspace{-0.4em}
\subsection{Focus+Context-based Image Exploration}
Cockburn et al.\cite{cockburn_review_2009} categorize interaction techniques to work at multiple levels of detail into focus+context (F+C), overview+detail (O+D), zooming, and cue-based views.
F+C minimizes the seam between views by displaying the focus within the context, O+D uses spatial separation, zooming temporal
separation~\cite{van_wijk_smooth_2003}, and cue-based methods selectively highlight or suppress items. Comparative studies show that F+C techniques are often preferred and allow for efficient and effective target acquisition~\cite{shoemaker_supporting_2007} and steering tasks~\cite{gutwin_fisheyes_2003} in multi-scale scenarios. A common F+C technique is the lens~\cite{bier_toolglass_nodate},
a generic see-through interface that lies between the application and the cursor. 
Tominski et al.~\cite{tominski_survey_2014,tominski_interactive_2017} present a conceptual pipeline for lensing consisting of selection (what data), the lens-function (filters, analysis), and a join operation with the underlying visualization (mapping, rendering). They further categorize into lens properties (shape, position, size, orientation), and into data tasks, e.g., (geo)spatial analysis. 
Different lenses to magnify, select, filter, color, and analyze image data were proposed:
Carpendale et al.~\cite{carpendale_extending_1997} present a categorization of 1-3D distortion techniques to magnify in 2D uniform grids. Focusing on lens-based selection, 
MoleView\cite{hurter_moleview_2011} selects spatial and attribute-related data ranges in spatial embeddings and Trapp et al. present a technique for filtering multi-layer GIS data for city planning~\cite{trapp_3d_2008}. Similarly, Vollmer et al. propose a lens to aggregate ROIs in a geospatial scene to reduce information overload~\cite{vollmer_hierarchical_2018}. 
Flowlens~\cite{gasteiger_flowlens_2011} features a lens for biomedical application: to minimize visual clutter and occlusions in cerebral aneurysms. 
There are a few tools in digital histopathology with lensing capabilities. Vitessce~\cite{gehlenborg_vitessce_2021}, positions linked views around an image viewer~\cite{manz_viv_2020} and includes a lens to show a predefined set of channels. However, by design, they do not focus on supporting a specific pathology process, nor does the lens support magnification, feature augmentation, comparison, or search.

\subsection{Handling and Visualization of ROI Annotations}
Different techniques exist to mark, visualize, and extract ROIs in images, but only a small subset is used in the digital pathology domain. QuPath~\cite{bankhead_qupath_2017}, an extensible software platform, allows annotating histology images with free form selection tools and more advanced selection options like pixel-based nearest neighbors and magic wand, extending from the clicked pixel to neighboring areas with a threshold. Visopharm's viewer~\cite{visopharm} similarly provides different geometric shapes to annotate images. In Orbit~\cite{stritt_orbit_2020} and Halo~\cite{indica_labs_halo_nodate} users can define inclusion and exclusion annotations, organize them in groups, and train a classifier.
Going beyond manual annotation, Quick-Annotator~\cite{miao_quick_nodate} leverages a deep-learning approach to search and suggests regions similar to a given example. Similarly, Ip and Varnsesh~\cite{ip_saliency-assisted_2011} 
narrow down and cull out ROIs of high conformity and allow users to interactively identify the exceptional ROIs that merit further attention on multiple scales. We incorporate these ideas but instead apply a fast neighborhood-based histogram search running on multiple image channels in real-time to guide the user to similar areas in the viewport.

When large images are annotated on different scales it becomes challenging to navigate in an increasingly cluttered space. Some features might even be too small to be identifiable at certain zoom levels. Scalable Insets~\cite{lekschas_pattern-driven_2020} is a cue-based technique that lays out regions of interest as magnified thumbnail views and clusters them by location and type. TrailMaps~\cite{zhao_trailmap_2013} proposes an algorithm to automatically create such insets (here bookmarks) based on user interaction and previously viewed locations. They also offer timeline- and category-based groupings for a better overview and faster navigation.
We choose a more familiar design to cater to the application domain needs and conventions but enhance our approach by supporting \textit{rich} annotations that store not only geometry but also linked single-cell data and descriptive statistics. Closely related to our approach is the work by Mindek et al.~\cite{mindek_managing_2014} that proposes annotations linked to contextual information so that they remain meaningful during the analysis and possible state changes. We extend this idea with overview, search, and restoring capabilities integrated into focus+context navigation in large-scale multivariate images.

\section{Background: Multiplex Tissue Imaging}
\label{sec:process}
\begin{figure}[t]
  \centering
  \includegraphics[width=\linewidth]{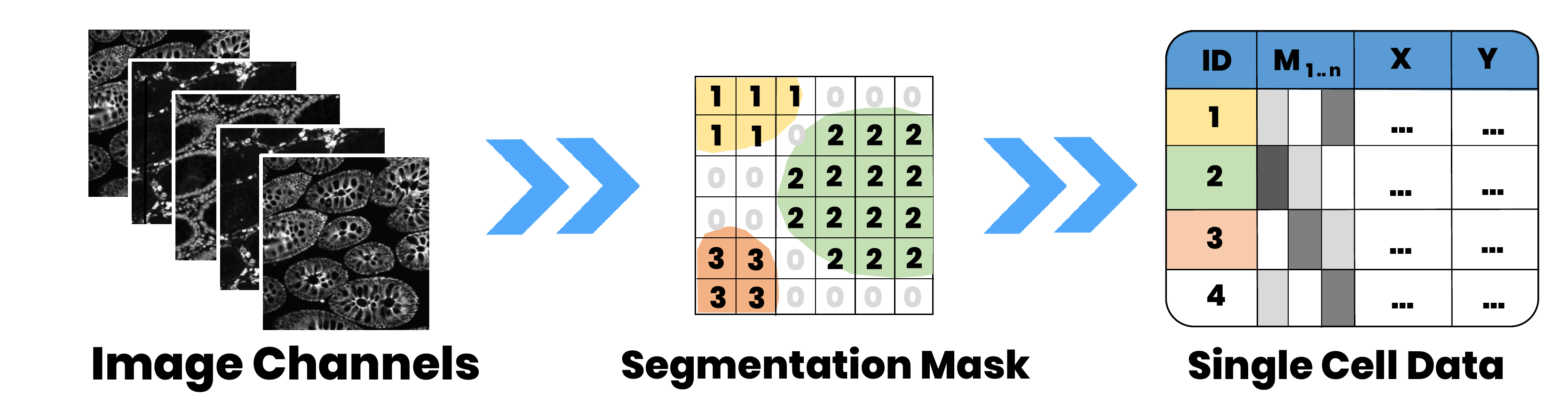}
    \vspace{-2em}
  \caption{Our histological tissue image data consists of a multi-channel image stack, a segmentation mask, and extracted tabular marker intensity values (arithmetic mean) for each cell. The tabular data is linked via cell ID and X,Y position.}
  \label{fig:data}
  \vspace{-2em}
\end{figure}

We analyze multiplexed tissue imaging data generated with CyCIF~\cite{lin_cyclic_2016} but our visualization approach can be applied to images acquired using other technologies such as  CODEX~\cite{goltsev_deep_2018}. Images are segmented and signal intensity is  measured at a single-cell level. Here we provide a brief overview of the process and data (Fig.~\ref{fig:data}).

\noindent
\textbf{Acquisition.}
\label{sec:ImagingWorkflow}

Multiplexed tissue imaging allows to analyze human tissue specimens obtained from patients for pathologic diagnosis. The approach used by the investigators, as described in our previous work~\cite{krueger_facetto_2020}), involves iterative immunofluorescence labeling with 3-4 antibodies to specific proteins followed by imaging with a high-resolution optical microscope in successive cycles. This results in 16-bit four-channel image datasets for up to 60 proteins of interest (60 images), 30k x 30k in resolution, and often greater than 100GB in size, allowing for extensive characterization and correlation of markers of interest in large tissue areas at sub-cellular resolution.

\noindent
\textbf{Processing.}
\label{sec:ImageWorkflowProcessing}
High-resolution optical microscopes have limited fields of view, so large samples are imaged using a series of individual fields which are then stitched together computationally to form a complete mosaic image using software such as  ASHLAR~\cite{muhlich_jeremy_et_al_labsyspharmashlar_2021, muhlich2021stitching}. A nonrigid (B-spline) method~\cite{klein_elastix_2010,marstal_simpleelastix_2016} is applied to register microscopy histology mosaics from different imaging processes~\cite{borovec_anhir_2020}, e.g., CyCIF and H\&E. 
CyCIF mosaics can be up to 50,000 pixels in each dimension and contain as many as 60 channels, each depicting a different marker. Mosaic images are then classified pixel-by-pixel to discriminate cells using, e.g., a random forest~\cite{sommer_ilastik_2011}, then individual cells are segmented~\cite{lsp_labsyspharms3segmenter_221}. 
Segmentation information is stored in 32-bit masks that define the cell ID for each pixel in a multi-channel image stack. 
Next, per-cell mean intensities are extracted for the $10^{6}$ or more individual cells in a specimen. The processing steps are combined in an end-to-end processing pipeline named MCMICRO~\cite{schapiro_mcmicro_2021}. The resulting 16bit multi-channel images ($\approx$100GB), 32bit segmentation ($\approx$5GB), and high-dimensional feature data ($\approx$2GB) are then ready for interactive analysis.

\noindent
\textbf{Terminology and Data Characteristics.}
Our datasets contain (1) a multi-channel tissue image stack with 1-60 channels in OME-TIFF format~\cite{noauthor_ome-tiff_nodate}, (2) a segmentation mask also in TIFF format, and (3) a table of extracted image features in CSV format (Fig.~\ref{fig:data}).
Each \emph{image channel} in the \textbf{multi-channel image stack} represents data from a distinct antibody stain and is stored as an image pyramid (in the OME-TIFF) for efficient multi-resolution access. These channels can result from different imaging processes (e.g., CyCIF and H\&E).
A \textbf{segmentation mask} labels individual \emph{cells} in each tissue specimen with a unique \emph{cell ID}. Similar to each image channel, the mask is stored in pyramid form.
A CSV file stores \textbf{single-cell features} (columns) for each cell (row). These features consist of extracted mean intensity values per image channel for that cell, x and y position of the cell in image space, and its cell ID.

\section{Domain Goals and Tasks}
\label{sec:goalsAndTasks}
\begin{figure*}[t]
  \centering
  \includegraphics[width=0.97\linewidth]{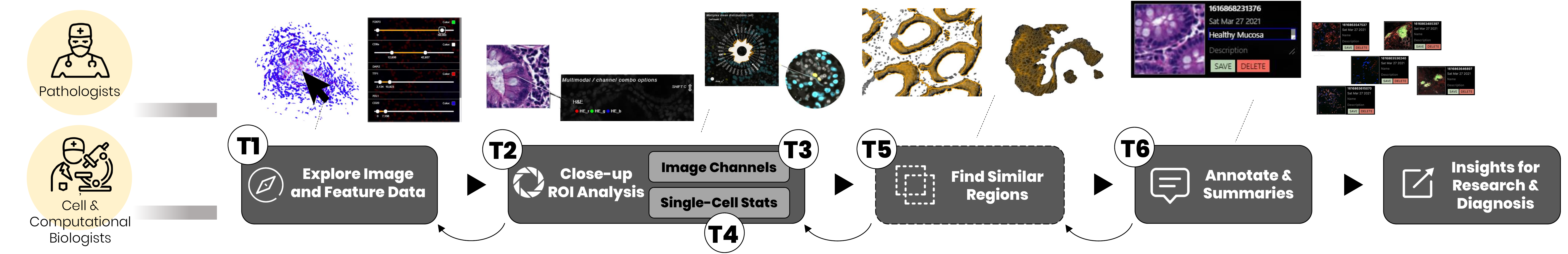}
    \vspace{-1em}
  \caption{The pathological workflow 
  starts with exploratory navigation in the image (T1). ROIs are magnified, measured, and analyzed (T2) by switching and combining image channels (T3) and investigating single-cell marker statistics (T4). Identified regions often appear in patterns across the image. Finding such similar regions (T5) can ease manual search. ROIs are then annotated (T6). These steps build an iterative process where annotations are refined, and further areas are explored. The ROIs are stored or exported to discuss with colleagues or for examination.}
  \vspace{-1em}
  \label{fig:workflow}
\end{figure*}

This project is rooted in a collaboration with physicians and researchers in the Department of
Pathology and the Laboratory of Systems Pharmacology (LSP) at Harvard Medical School. Four experts in the domain of digital histopathology participated in the project.
The team consists of two pathologists, two computational biologists, and four computer scientists. The \textbf{overall goal} of our collaborators is to characterize the features of tumors including cell types \& states, their interactions, and their morphological organization in the tumor microenvironment.

\noindent\textbf{Pathologists} are physicians who diagnose diseases by analyzing samples acquired from patients. Anatomic pathologists specialize in the gross and microscopic examination of tissue specimens. They characterize cell and tissue morphology using light microscopy, and molecular features using immunohistochemistry and immunofluorescence. 
The pathologists involved in this project engage in research and have expertise in imaging, computational biology, and defining the role of diverse cell states in shaping and regulating the tumor microenvironment.

\noindent\textbf{Computational and Cell Biologists} complement expertise in biomedical science with skills in technical fields including mathematics, computer science, and physics. Multiplexed immunofluorescence experiments involve collection of primary imaging data which is used for a wide variety of complex computational tasks including image registration and segmentation, and extraction of numerical feature data, as well as downstream analyses of cell states, spatial statistics, and other phenotypes. Biologists interpret aspects of cell morphology and marker expression, but pathologists complement these analyses with greater depth of experience with human tissue morphology and disease states.

\noindent\textbf{Visualization Experts.} By contrast, the computer scientists provide expertise in 
visualization and visual data analysis. They work in close collaboration with the aforementioned investigators to provide novel analytics prototypes that perform a variety of analysis tasks and can be integrated into research studies and laboratory IT infrastructure. To understand domain goals, the visualization experts in this study participated in weekly meetings focused on image processing, biomedical topics, and on iterative goal-and-task analyses for the proposed approach. The collaboration with the LSP started in 2018 with Facetto~\cite{krueger_facetto_2020}.

\smallskip
\noindent In Fall 2020, this team began working together 
to develop advanced tools that cater to visual exploration, inspection, and annotation.
We followed the design
study methodology by Sedlmair et al.~\cite{sedlmair_design_2012}.
This methodology describes a setting in which visualization experts analyze a domain-specific problem, design a visualization approach to solving it, validate their design, and reflect on lessons learned.

\subsection{Tasks and Challenges}
In weekly sessions, 
we identified a workflow (Fig.~\ref{fig:workflow}) of consecutive tasks \textbf{(T1-T6)} leading from image exploration and close-up inspection of regions to annotation and extraction of patterns.

\noindent\textbf{T1. Explore Multimodal, Highplexed Image and Feature Data in Combined Setting:}
A pivotal task is rapid navigation and
visualization of multi-channel images. Pathologists normally operate by moving slides physically on a microscope stage and switching between view magnifications levels. They
depend on a seamless visual experience to diagnose diseases or conditions. \textit{Challenge:} Image analysis must not only provide seamless pan \& zoom, but
also switching between channels of different image modalities. 
Existing solutions do not scale beyond 4 to 5 channels. They also lack on-demand rendering, blending of multiple channels, and ways to highlight and recall ROIs.

\noindent\textbf{T2. Close-up ROI Analysis:}
Once a region of interest is found in the tissue specimen, experts focus and zoom in on the area for close-up inspection and measurement, without losing the spatial context of e.g., a tumor region's surrounding immune cells. \textit{Challenge:} In addition to interactive rendering of different resolution levels in a combined space, experts need to focus and measure without losing proportions and larger context. Panning and zooming between overview+detail and individual marker channels requires a large amount of mental effort as either context or details are lost~\cite{lekschas_pattern-driven_2020}.

\noindent\textbf{T3. Regional Comparison of Image Markers:} For a region in focus, cell biologists need to relate and compare between different marker expressions (e.g., DNA, CD45, Keratin) of different image modalities (CyCIF~\cite{lin_highly_nodate, lin_cyclic_2016}, H\&E~\cite{titford_long_2005}, CODEX~\cite{goltsev_deep_2018}, mxIF~\cite{gerdes_highly_2013}, etc.), encoded in individual image channels. \textit{Challenge:} 
Whole-slide switching between channels can lead to losing focus and change blindness due to different morphological structures.

\noindent\textbf{T4. Relate to Spatially Referenced Single-Cell Expressions:}
Besides looking at the raw image, experts analyze extracted singe-cell marker values and their spatial statistics in (A) image and (B) high-dimensional feature space. Of special interest is cell density in the tissue, counts and spatial arrangements of cell-types, and distributions of marker intensities. For each of these descriptive statistics, it is important to relate regional phenomena to statistics of the whole image. \textit{Challenge:}  Providing complex spatially referenced information
in proximity while dealing with a dense cellular image space and catering to highly specific domain conventions.

\noindent\textbf{T5. Find Similar Regions:}
Analyzing a whole slide image is time-consuming. Often the cancer micro-environment consists of repetitive patterns of cell-cell interactions and morphological structures across channels that pathologists annotate and compare to each other. \textit{Challenge:} Finding and guiding users to such structures in an interactive fashion on different spatial scales and across image dimensions.

\noindent\textbf{T6. ROI Annotation and Summaries:}
A common pathology task concerns the manual delineation of tumor mass and other structures on the digitized tissue slide, known as region annotation. These annotated regions need to be extracted, semantically grouped, and summarized in a structured way for collaboration and examination.
\textit{Challenge:} The annotation process must be integrated seamlessly with the analysis so that experts can extract, group, and refine patterns along the way.

\section{Approach}
We used the tasks of Section 4 to guide the design and implementation of Scope2Screen, playing the translator role put forth in the design study methodology by Sedlmair et al.~\cite{sedlmair_design_2012}. 
Fig.~\ref{fig:teaser} offers an overview of the user interface. After importing the data, analysts can explore (Fig.~\ref{fig:teaser}) the image by activating a set of different channels (A) with distinct color configurations and adjustable intensity ranges. The selections are then rendered in a combined view (B) for interactive panning and zooming \textbf{(T1)}. Using a virtual lens (C), users can focus, magnify, and measure regions of interest for close-up analysis \textbf{(T2)}.
By toggling image channel combinations inside and outside of the lenses, one can regionally compare different combinations of marker expressions \textbf{(T3)}. Other lens filters link to underlying single-cell data, offering descriptive statistics about marker distributions and cell counts and types \textbf{(T4)}. Using a search-by-example approach, the tool guides users to regions with similar patterns as those in scope \textbf{(T5)}. To save and extract a region, analysts can take snapshots that save the ROI together with relevant notes (Fig.~\ref{fig:teaser} D), current settings, and interior single-cell data for the region \textbf{(T6)}. Annotated areas can be filtered and exported to share with collaborators and to recall for further examination.
In the following sections, we introduce the corresponding techniques and features \textbf{(F1-F6)} and discuss design decisions to enable these tasks.

\begin{figure*}[t]
  \centering
  \includegraphics[width=0.90\linewidth]{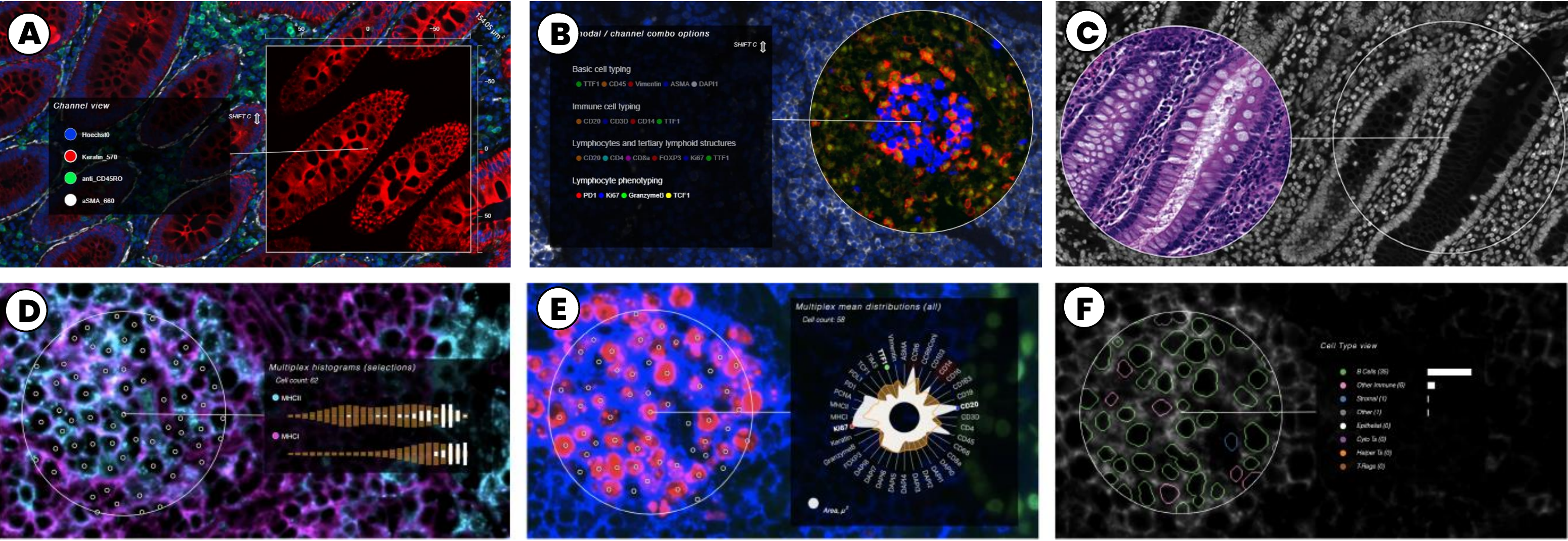}
    \vspace{-0.8em}
  \caption{Top: Settings for channel analysis: (A) Single channel option, out of three in the context. (B)  Multi-channel lens. (C) Split-screen lens enabling juxtaposed comparison of the same area with different multi-channel settings (here CyCIF-DNA and H\&E-RGB). Bottom: Feature augmentation: (D) Single-cell histograms for detailed vertical comparison of selected cell marker distribution (channel-based rendering); (E) Radial single-cell plot a for compact summary of cell marker distribution; (F) Segmentation, cell types and counts showing classification results.}
  \label{fig:featureAugmentation}
  \vspace{-1em}
\end{figure*}
\begin{figure}[t]
\begin{subfigure}{.33\linewidth}
  \centering
  \includegraphics[width=0.95\linewidth]{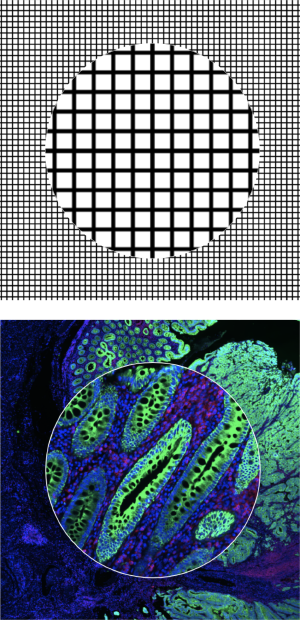}
  \caption{Magnification}
  \label{fig:sfig1}
\end{subfigure}%
\begin{subfigure}{.33\linewidth}
  \centering
  \includegraphics[width=0.95\linewidth]{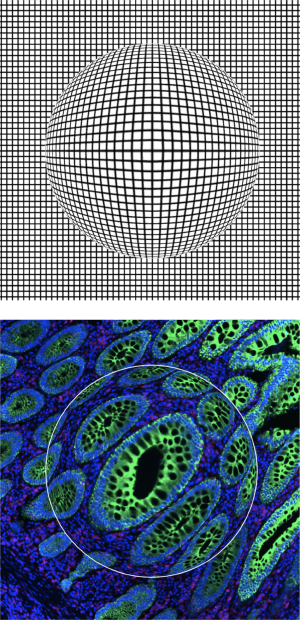}
  \caption{Fisheye}
  \label{fig:sfig2}
\end{subfigure}
\begin{subfigure}{.33\linewidth}
  \centering
  \includegraphics[width=0.95\linewidth]{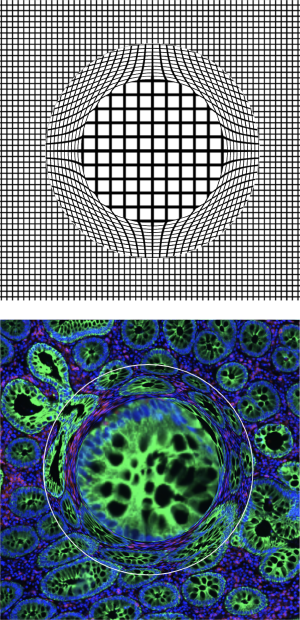}
  \caption{Plateau}
  \label{fig:sfig2}
\end{subfigure}
\vspace{-1em}
\caption{Magnification options: (A) normal magnifier; (B) fisheye, introducing distortion with an interpolated spherical shape; (C) plateau with 75\% preserved resolution and 25\% compressed interpolation. High-resolution image quality within the zoom area is achieved by accessing image data from more detailed layers in the image pyramid.}
\label{fig:magnifiers}
\vspace{-2.0em}
\end{figure}
\subsection{Scalable Image Exploration (F1)}
\label{sec:imageExploration}
We designed Scope2Screen for interactive exploration and multi-scale visualization~\textbf{(T1)} of resection specimens
(Sec.~\ref{sec:process}). To make the viewer scalable for high-resolution data, we decided to leverage image pyramids~\cite{in_e_1984} to load only sections of the image for a given viewport and zoom level.
To enable flexible exploration of image channels and to visualize data in the viewport, the viewer operates in two multi-resolution rendering modes: \textbf{channel-based} and \textbf{cell-based}. Channel-based rendering maps intensity values of selected image channels to color. It then computes a mixed color value for each pixel in the viewport.
Cell-based rendering leverages a layered segmentation mask that indexes each pixel to a cell. This way, each cell can be colored individually to visually express selections, cell types, etc. Both rendering modes operate at interactive rates, allowing users to pan \& zoom, select regions, and to color and mix channels in real-time. These rendering modes cater to our experts' needs to analyze on tissue and single-cell level.
Sec.~\ref{sec:implementation} gives details on the implementation of our system.

\subsection{Lensing Features for Multivariate Image Data}
\label{sec:Lensing}
Conveniently, the principal technology of the microscope, the lens, is highly adaptable. To enable close-up analysis \textbf{(T2)} of ROIs and to connect the optical and digital experiences for users, we introduced a digital lens designed to imitate the familiar experience of inspecting through an eyepiece. 
To support the requirements of our collaborators, we equip the lens with features (Fig.~\ref{fig:featureAugmentation},\ref{fig:magnifiers}), ranging from magnification and channel filters \textbf{(F3)} to descriptive single-cell statistics \textbf{(F4)}. 

\subsubsection{Magnification and Measuring (F2)}

Our users rely on being able to toggle quickly between high and low magnification powers as part of an established workflow for considering a region of interest up-close, as a localized arrangement of cells, and in-context, as part of a larger tissue sample \textbf{(T2)}. Within the virtual viewer, this zoom-level interchange can be challenging to control. Constraining magnification to the lens's boundary (Fig.~\ref{fig:magnifiers}) while maintaining the contextual overview then becomes a convenient strategy for handling simultaneous focal analysis.
Because the magnifying lens (Fig.~\ref{fig:magnifiers} A), when active, occludes part of the image space, we experimented with a common spatial manipulation to create a faux-spherical representation: the fisheye (B). However, distortion is a troubling approach for experts who make evaluations based on morphology, leading us to introduce a hybrid plateau model (C) that maintains the original composition within the central area of the lens using a standard zoom and only compresses the periphery to seamlessly transfer into the context without occlusion.
The scale of an ROI (in microns) is closely tied to magnification interpretability. Users consider area-based standards for clinical validation as part of their inspection methodology. Visible axes around the lens allow for a quick understanding of scale (Fig. ~\ref{fig:teaser} C). Additionally, this embedded conversion capability between digital and physical units is a useful tool for extended functionalities that emulate related user tasks (e.g., cell prevalence counting and density analysis).

\subsubsection{Channel Analysis (F3)}
\label{sec:ChannelAnaysis}

To address regional comparison of channels \textbf{(T3)} that represent data from the same modality (e.g., CyCIF), other modalities (e.g., H\&E~\cite{titford_long_2005}), and across different planes (e.g., slide sectioning), we iterated over different designs and finally settled on the following features.

The first lens feature allows users to quickly isolate each selected channel individually for improved views of a distinct channel (Fig.~\ref{fig:featureAugmentation} A) in focus while keeping the multi-channel setting in the context.
The second feature combines multiple channels in the scope (Fig.~\ref{fig:featureAugmentation} B, Fig. \ref{fig:multimodal_lung}) while the context can keep a distinct setting (Fig.~\ref{fig:teaser}). We specifically designed this setting for semantically dependent channels such as RGB images for H\&E staining. It also addresses our experts' needs to analyze spatial relations of a set of independent channels, for example, from specific immune markers in a region of interest, while keeping globally a different set of more general cancer and stromal markers. Addressing early feedback from our experts, we added the capability to equip the lens with multiple sets of such channel combinations (Fig.~\ref{fig:multimodal_lung}) in advance or add them during analysis. These sets can be quickly toggled during exploration to investigate different biological questions, e.g., focusing on immune reactions, or tissue architecture.
To overcome occlusion, we chose to offer two solutions leveraging temporal and spatial separation. Firstly, we introduce adjustable interpolation controls allowing to \textit{blend} seamlessly from the overlaying lens-image to the underlying global channel combination. This transition helps to visually align and keep track of often very different structures in different channel combinations.
To further reduce change blindness, a split-screen lens juxtaposes a second lens instance in proximity, which displays (copies) the occluded part in the original global channel setting (Fig.~\ref{fig:featureAugmentation} C). This allows for side-by-side comparison.

\begin{figure*}[t]
  \centering
  \includegraphics[width=1.0\linewidth]{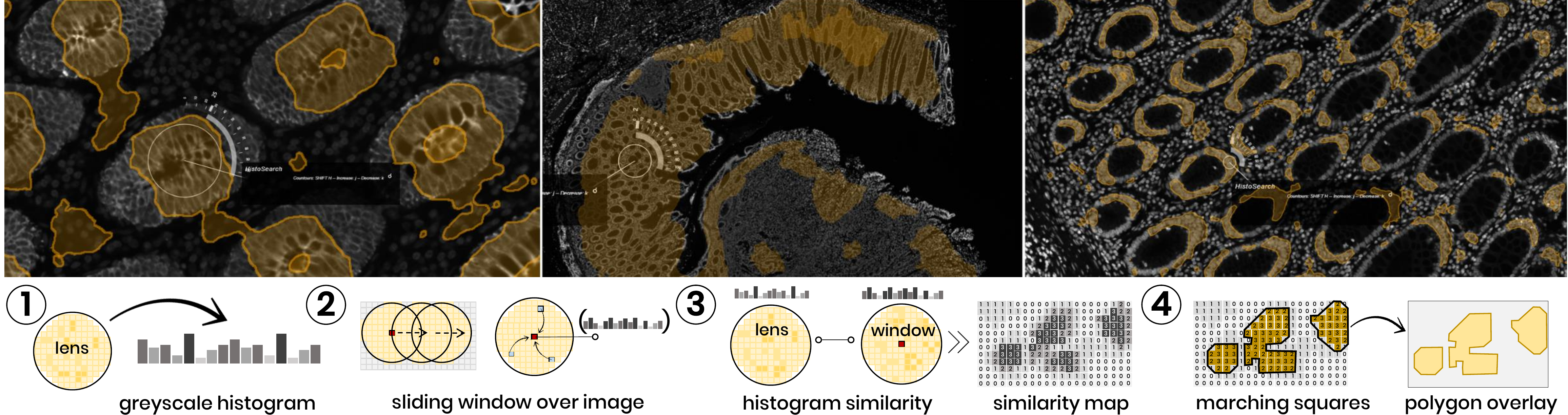}
    \vspace{-1.50em}
  \caption{HistoSearch allows to find regions similar to those covered by the lens, taking into account activated channels. Top: HistoSearch is applied at different scales to find  mucosal regions.  The search works in two settings, for the current viewport (computation time $\approx$ 1 second for Full HD) and for the whole image in the highest resolution. Bottom: The spatial histogram similarity search consists of four steps (Sec.~\ref{sec:similarRegions} for details). 
  }
  \label{fig:histoSearchResults}
  \vspace{-2.0em}
\end{figure*}

\subsubsection{Feature Augmentation (F4)}
\label{sec:featureAugmentation}
Our multiplex image data (Sec.~\ref{sec:process}), often comprises up to 60 channels. While three channels can guarantee non-overlapping (RGB) colors, adding even more channels makes visual decoding for analysts mentally challenging and increasingly inaccurate. Additionally, color encodings of quantitative data are often hard to gauge~\cite{mackinlay_automating_1986}. Instead, to enable quantitative analysis of selected regions,
we chose to augment the image space with descriptive statistics of the extracted single-cell data~\textbf{(T4)} using more abstract visual encodings (Fig.~\ref{fig:featureAugmentation} D-F). We developed three task-tailored lens settings showing marker distributions, density reports, and cell types and counts. With every update of the lens position on screen, the back-end queries the in-memory CSV table (Sec.~\ref{sec:process}) for cells in the lens's area and returns cell Id values along with the requested statistics, which are then processed and rendered into different charts.
To speed this up, we execute spatial range queries with a ball-tree index structure~\cite{dolatshah_ball-tree_2015}. This yields a run time complexity of $\mathcal{O}(n\log{}n)$.

\noindent\textbf{Single-Cell Histograms.} To analyze marker distribution in a selected region, we compute binned histograms of the single-cell aggregates (see CSV table, Sec.~\ref{sec:process}) for selected channels. We present these histograms (Fig.~\ref{fig:featureAugmentation} D) in proximity to the focus area for quick look-ups. We decided to arrange the histograms in a vertical layout to ease comparison. According to our domain collaborators, absolute comparison of individual markers is statistically not meaningful as the signal-to-noise ratio changes per cycle and stain in the imaging process (Sec.~\ref{sec:process}). Instead, they are interested in a relative comparison of the distributions. We use a $log_{2}$ transformation to make these skewed marker distributions comparable, followed by a cut-off of the 1st and 99th percentile to remove outliers. Fig.~\ref{fig:teaser} (C) shows the lens operating in this setting. To further analyze where cells lie in that spectrum, we chose to offer a brush functionality. The user can filter a range (min-max) in the histogram, which highlights cells in the lens matching the updated single-cell marker values.

\noindent\textbf{Radial Chart.}
Additionally to histograms, we provide an overview of all markers by arranging their mean values in a radial layout in proximity to the lens (Fig.~\ref{fig:featureAugmentation} E). This decision allows for a more compact representation of the multivariate single-cell data.
When we first showed this to our collaborators, they missed a reference to compare the region of interest to global image statistics. In a second design iteration, we thus encoded arithmetic means for the whole tissue. The histograms show these whole-slide references as orange bins (background) and the radial plot encodes the information as a polyline.

\noindent\textbf{Cell Types and Counts.}
Our collaborators want to validate results of cell-type classification and clustering~\cite{liu_comparison_2019,campanella_clinical-grade_2019} and set them into spatial context to expressions in the image channels.
We color-code cell boundaries by detected cell types, using the cell-based rendering mode. This mode utilizes the segmentation mask to retrieve which pixel is linked to which cellId (Sec.~\ref{sec:imageExploration}). The colored boundaries (Fig.~\ref{fig:featureAugmentation}, right) allow to display the data at its spatial image position and still see the channel colors. As classification often strongly depends on the expressions in a few channels, users can pick matching channel colors and check if the pattern spatially aligns with the cell types.
To ease quantification in the localized region, we also provide an ordered list of cell types and counts near the lens. Counts are encoded as bars that dynamically adapt while hovering over the data. Users can switch between two orderings: Locked cell type positions allow to observe increasing or decreasing presence of specific types; ranking by count is preferred to spot the most prominent cell types in the focus region.

\subsection{Finding Similar Regions (F5)}
\label{sec:similarRegions}

\begin{figure*}[h]
  \centering
  \includegraphics[width=0.85\linewidth]{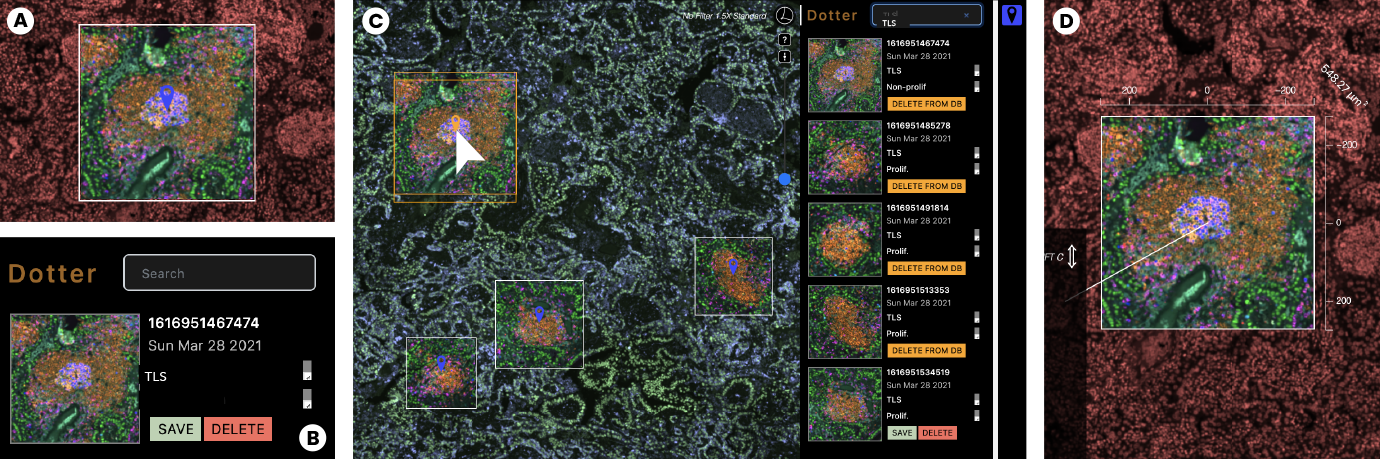}
    \vspace{-0.5em}
  \caption{The rich snapshot and annotation process. (A) During close-up analysis, the user focuses on an ROI and takes a snapshot. (B) The snapshot is annotated with title and description. (C) The Dotter panel links snapshots to the image space (left). Lens-settings such as channel combination and colors are preserved. (D) Annotated regions can be \textit{reactivated} as lenses to explore further or fine-tune.}
  \vspace{-1em}
  \label{fig:annotaton}
\end{figure*}

Once identified, a region of interest often serves as a reference point to find similar areas within the image \textbf{(T5)}. Examples for repetitive patterns are, e.g., tumor-immune boundaries or germinal centers where mature B cells proliferate, differentiate, and mutate their antibody genes. 
To find similar regions, we chose to provide a method operating directly on the image to align as close as possible with the visual perception (see Fig.~\ref{fig:histoSearchResults}).
We consider regions similar if they have a similar intensity distribution. To compare a region to all possible areas in an image, we employ a sliding-window strategy that compares histograms of regional marker intensity distributions across the image. To trigger the search, the HistoSearch lens can be placed over the pattern of interest in the image. HistoSearch is equipped with a slider to set a contour threshold, allowing for fine-tuning of what's considered similar. 
The applied integral histogram method~\cite{porikli_integral_2005,perreault_median_2007} makes it possible to employ even an exhaustive search process in real-time. We adapt and extend a Python implementation~\cite{noauthor_sliding_nodate}. Our method works in four steps (Fig.~\ref{fig:histoSearchResults}, 1-4):
\smallskip

\noindent\textbf{Step 1:} A box- or circle-shaped region (the lens area) is extracted, and a histogram of its greyscale values is computed (Fig.~\ref{fig:histoSearchResults}, Step 1).

\smallskip

\noindent\textbf{Step 2:} For each pixel in the whole channel image, a histogram of the greyscale values surrounding (lens radius) the pixel is computed. Semantically, this means that we take into account spatial neighborhood information and not simply compare pixel by pixel (Fig.~\ref{fig:histoSearchResults}, Step 2). 

\smallskip

\noindent\textbf{Step 3:} The local histogram for the region surrounding each pixel in the image is then compared to that of the lens using Chi-square distance. We apply Porikli's integral histogram~\cite{porikli_integral_2005}, which recursively propagates histogram bins of previously visited image areas using values from neighboring data points instead of repetitively executing the full histogram computation. This is then compared to the lens histogram using Chi-square distance (see Eq.~\ref{eq:chi}) for two arrays $X$ and $Y$ with $N$ dimensions across all channels $C$. This leads to a similarity map with a similarity value for each pixel in the image  (Fig.~\ref{fig:histoSearchResults}, Step 3).

    \vspace{-0.5em}
\begin{equation}
\label{eq:chi}
    \frac{1}{c}\sum_{c=1}^{C}{\sum_{i=1}^{N}{((x_i-y_i)^2 / (x_i+y_i))}}
\end{equation}
\noindent\textbf{Step 4:} We use marching squares~\cite{perreault_median_2007} to detect contours in the similarity map and extract these contours as polygons that we render on screen (Fig.~\ref{fig:histoSearchResults}, Step 4).

\smallskip
\noindent Step 2 and 3 can be computed for multiple channels. To not lose information, we compute each channel's similarities separately and then aggregate them to a combined similarity map.
Similar to our multi-resolution rendering strategy (Sec.~\ref{sec:imageExploration}), we execute the histogram search algorithm on the fly on the respective TIFF-file layer (Sec.~\ref{sec:process}) that aligns with the current zoom level. Fully zoomed out, the aggregation level is higher; while zoomed in, we process the highest resolution but for a smaller image area. The approach can also be employed in full resolution to the whole image (Sec.~\ref{sec:ROIsummaries}).

\subsection{Descriptive ROI Summaries (F6)}
\label{sec:ROIsummaries}

When using a light microscope, a pathologist often `dots' whole slides with a pen to indicate ROIs for later examination. We introduced a digital ``dotter'' to represent this approach but with extended support for annotation \textbf{(T6)} during exploration (Fig.~\ref{fig:annotaton}). The lens functions as a camera lens that can snapshots whenever an ROI is in focus. 

Our collaborators currently narrate image annotations to data stories for examination, teaching and outreach of their research~\cite{rashid_interpretative_2020}. The link to analysis results is often lost as annotations must be manually recreated in the used tools.
To maintain analytic provenance, we developed a novel \textit{rich snapshot} method that not only saves the image area under the lens but also stores all associated data: a list of active image channels in focus and in the peripheral context, channel colors, and range settings, the linked single-cell data in scope, and textual annotation such as title and a more detailed description pathologists and cell biologists can add.

These rich snapshots are available as thumbnails in the Dotter panel (Fig.~\ref{fig:annotaton} C) and are interactively linked to overlays within the viewport. Inside the panel, the user can save, delete and load ROIs from a database. Names and descriptions can be edited and referenced as tags for filtering results. It is important to our users to be able to quickly recall and restore the snapshots for further analysis and fine tuning, but also to return to ROIs in their original context. Thus, by clicking on thumbnails, the viewer navigates back to the coordinates and zoom level of the overlay. By clicking on the overlay's marker icon, we fully restore the lens, along with its global view setting, i.e. active channels, range and color mappings for the context (Fig.~\ref{fig:annotaton} D). This workflow facilitates hand-offs between colleagues who benefit from shared evaluation.

To extend the search for an ROI in the Dotter panel, users can use HistoSearch (Sec.~\ref{sec:similarRegions}) to find regions alike in the whole image space (other than the viewport during interactive analysis). We render these annotations as image-overlays but as soon as the user picks up a region, we restore the lens, and the user can alter the region as needed.

\section{Architecture and Implementation}
\label{sec:implementation}

Scope2Screen is an open-source web-application (available here:~\cite{scope2screen}) with a back-end Python server built on Flask and a Javascript frontend. The server's restful API allows to retrieve image and feature data and to steer analytics and is easily extendable with new methods and corresponding API endpoints. The frontend is built using Bootstrap, D3.js, and OpenSeadragon (OSD)~\cite{noauthor_openseadragon_nodate}, a web-based zoomable image viewer, which we extend significantly.

We developed a lensing library that supports a subset of the interactive features of Scope2Screen as a generic plugin~\cite{jared_lensing} for OSD. The library relies on a hidden viewer that provides access to both in-view and out-of-view image data to support controlled rendering within the lens, along with other complementary features. Our application utilizes and extended lensing's logic with additional features (Sec.~\ref{sec:Lensing}).

Scope2Screen also builds on Facetto~\cite{krueger_facetto_2020} but makes improvements to the architecture. Instead of preprocessing OME-TIFF~\cite{noauthor_ome-tiff_nodate} image stacks to DeepZoom~\cite{microsoft_silverlight_2021} we now read chunks (cropped 2D arrays of the respective layer in the image channel/mask) on-the-fly from layered OME-TIFFs to render it in the viewport at multiple resolution levels, depending on the current zoom setting. We rely on Zarr~\cite{alistair_miles_zarr-developerszarr-python_2020}, a format for the storage and handling of chunked, compressed, N-dimensional arrays. The client-side rendering is hardware accelerated. It relies on WebGL~\cite{noauthor_webgl_nodate} shaders from Minerva~\cite{hoffer_minerva_2020,rashid_interpretative_2020} and supports Facetto's channel and cell-based rendering modes (Sec.~\ref{sec:imageExploration}) for both the lens (focus) and the whole viewport (context). To access and filter the linked single cell feature (CSV) data more dynamically and at scale, we moved data processing and ball-tree~\cite{dolatshah_ball-tree_2015} indexed querying (Section~\ref{sec:featureAugmentation}) to the back-end so that the client only loads requested pieces (in lens or viewport), aligning with our multi-resolution rendering strategy.

\section{Use Cases}
\label{sec:useCases}

We applied Scope2Screen to study two cancer datasets that we acquired from sections of lung and colon cancer using CyCIF~\cite{lin_cyclic_2016}.
Immediately adjacent sections were H\&E stained~\cite{titford_long_2005} and used to evaluate tissue morphology. We carried out the analysis together with our collaborators over zoom using a Pair-Analytics approach~\cite{arias-hernandez_pair_2011}; we steered the tool guided by the domain collaborators. This method is advantageous because it pairs Subject Matter Experts (SME) with expertise in a domain with Visual Analytics Experts (VAE) who have the technical expertise in the operation of VA tools but lack contextual knowledge.

\subsection{Use Case 1: Colorectal Cancer}
\label{sec:UseCase1}

Tumors are complex ecosystems of numerous cell types arranged into recurrent 3D structures. However, the patterning of specific tumor and immune cell-states is poorly understood due to the difficulty of mapping high-dimensional data onto large tissue sections. In use case 1, two pathologists and one cell biologist analyzed a human colorectal carcinoma (CRC1) from the Human Tumor Atlas Network (HTAN) (PMID 32302568) to explore tumor and immune cell interactions.

\noindent\textbf{Data:} 
CRC1 is a human colorectal adenocarcinoma that was serially sectioned at 5um intervals into 106 sections. 24-marker CyCIF was performed on every 5th section. Every adjacent section was stained with H\&E. CyCIF and H\&E images were registered and stitched, and single-cell segmentation and fluorescence intensity quantification were performed using MCMICRO~\cite{schapiro_mcmicro_2021}. Cell types were defined by expression of lineage- \& state-specific markers. Here, we analyze one CyCIF and an adjacent H\&E image. The data included 40 CyCIF channels and 3 (RGB) H\&E channels encompassing 1.28 mio cells in a field  26,139 x 27,120 pixels (8,495 x 8,814 microns) in size (87.66 GB).

\begin{figure}[t]
  \centering
  \includegraphics[width=0.90\linewidth]{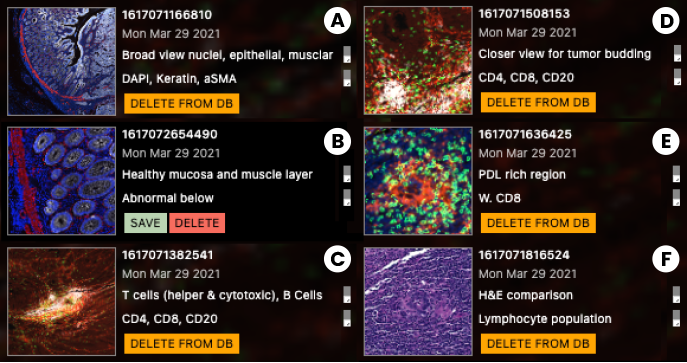}
      \vspace{-0.75em}
  \caption{
  Use Case 1. Rich snapshots capture ROIs and important insights: (A) Broad population; (B) Healthy tissue; (C) Immune cell rich; (D) Tumor budding ; (E) Tumor suppression; (F) H\&E - lymphocyte.}
  \label{fig:useCase1}
  \vspace{-2em}
\end{figure}

\noindent\textbf{Analysis: }
We began the analysis with a low magnification overview of the CyCIF images using DAPI (blue), keratin (white), and aSMA (red) channels in the whole viewport. In combination, these channels illuminated pathologically-relevant structures of the tissue, including the morphology of nuclei (DAPI), abnormal epithelial cells (keratin), and muscular layers (aSMA) (Fig.~\ref{fig:useCase1} A, B). A second channel defined immune populations including CD4+ helper T cells (red), CD8+ cytotoxic T cells (green), and CD20+ B cells (white) to analyze immune interactions with tumor and adjacent normal regions (Fig.~\ref{fig:useCase1} C).

For each marker, we defined an upper and lower color mapping range using the channel range sliders (Fig.~\ref{fig:teaser} A). The low-magnification view revealed a small region of tumor budding cells ($\le$ 1mm$^2$), a phenomenon in which infiltrating single tumor cells or small clusters of cells ($\le$ 4) appear to ``bud'' from the tumor-stromal interface at the invasive margin, correlating with poor clinical outcomes (Fig.~\ref{fig:useCase1} D). We used the standard lens magnifier to focus analysis on the budding region while maintaining spatial context.

To further explore spatial patterns of marker expression, we activated the “single-cell radial chart”. It provides a dynamic display of the mean single-cell expression levels of all CyCIF markers within the lens and the global mean of the markers across the entire image for comparison. This enabled the experts to see that tumor and immune cells in the several regions, including the budding region, were positive for PD-L1, a protein that suppresses the activity of cytotoxic CD8 T cells, which is often clinically targeted by 
antibody therapies. To capture images and associated single-cell data of these ROIs for subsequent review of immune interactions and tumor features, we took  snapshots and annotated the areas 'PD-L1 rich region' (Fig.~\ref{fig:useCase1} E) for later analysis.

We next used the split-screen lens (Fig.~\ref{fig:featureAugmentation} C) to view H\&E and CyCIF images side-by-side. Using this tool, pathologists validated the alignment of H\&E and CyCIF channels acquired from adjacent sections to compare histologic and molecular features. They identified areas in the H\&E images with marked chronic inflammation (lymphocytes and macrophages) in the peri-tumoral stroma for further evaluation. Direct comparison of the H\&E and CyCIF data in these tissue regions using the split-screen lens allowed the pathologists to characterize lymphocyte aggregates with predominantly cytotoxic CD8+ T cells in peri-tumoral stroma  (Fig.~\ref{fig:useCase1} E), as well as clusters of CD20+ B cells and CD4+ helper T cells forming `tertiary lymphoid structures' (TLS). Direct comparison of H\&E and CyCIF data in this manner allowed for targeted molecular characterization of immune populations such as lymphocytes which is not possible with H\&E alone (Fig.~\ref{fig:useCase1} F). To compare the marker intensity value distributions more precisely, we switched to the `lens histogram' (Fig.~\ref{fig:featureAugmentation} A) and compared the results with the `cell-type' lens within each region to assess marker expression with the results of the automated cell type classifier.

Based on review of the images, the pathologists used the Dotter’s `snapshot' function to annotate three immunologically distinct regions (described above), three tumor regions with distinctive histomorphology (glandular, solid, and mucinous regions), and adjacent normal colonic mucosa. Using the `similarity search' on the normal mucosal region (Fig.~\ref{fig:histoSearchResults}, top), we identified areas with similar histologic features, confirmed by pathologist inspection, embedded within the tumor mass which were not readily apparent on initial low-magnification review of the H\&E images, validating the utility of the search method. We saved these findings to our database for subsequent retrieval.
\begin{figure*}[t]
  \centering
  \includegraphics[width=0.75\linewidth]{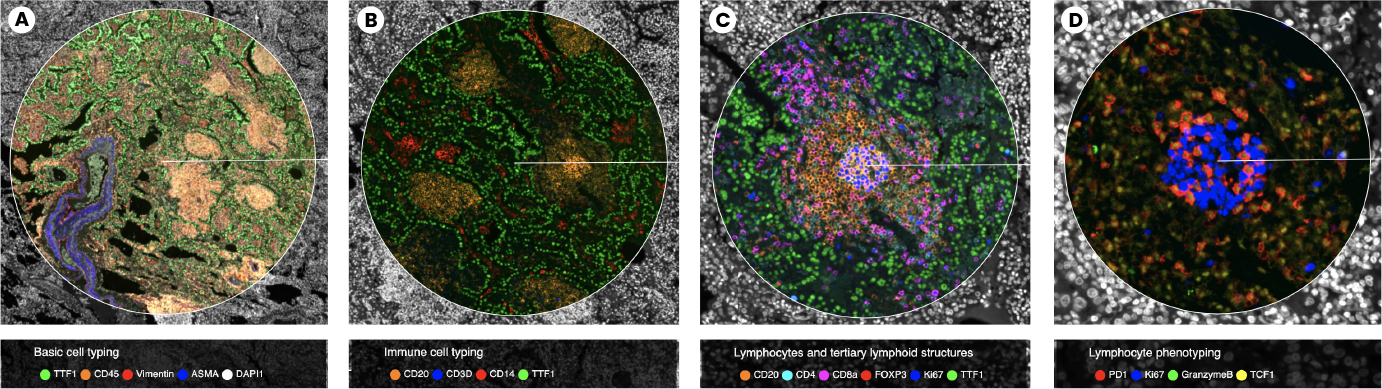}
    \vspace{-0.5em}
  \caption{Use Case 2, Multi-channel lenses in 4 settings: (A) `Basic Cell Typing' shows tissue composition - stromal, immune, and cancer cells. The dense structure is a result of tumor growth in the lung; (B) `Immune Cell Typing' distinguishes between immune and non-immune cells for a broad overview of immune regions (orange); (C) `Lymphocytes and TLS' combines CD-channels reveal distinct immune types, e.g., cytotoxic T cells attacking the cancer; (D) `Lymphocyte Phenotyping' for finer distinction, showing proliferating B-cells for antibody production (in blue).}
  \label{fig:multimodal_lung}
  \vspace{-1.5em}
\end{figure*}

\noindent\textbf{Feedback:}
Although we worked with our collaborators in weekly sessions over several months, we received additional comments on design improvements and future features within the 2-hour analysis session.
They found the normal lens magnification the most useful. The fisheye lens was problematic for review of tissue images due to distortion of cell morphology and tissue architecture, which may complicate the interpretation of important pathologic features such as nuclear shape. This aligns with earlier feedback motivating the design of the plateau lens, which they confirmed as a helpful improvement.
One pathologist also suggested to equip the lens with different predefined filter settings to be able to quickly toggle for different analysis tasks.
While both the radial chart and histograms were well received, they asked to highlight under- or over-represented marker values more prominently. The histograms were described as easier to read and should be extended to represent non-active image channels as well. They also asked for a) a heatmap that color codes marker correlation, and b) the ability to define channel combinations inside the lens on-the-fly, not only in pre-configuration. 
A recurring piece of feedback during the session was to add more descriptive statistics such as bin sizes in the histogram, ratios of cell populations, and ways to precisely define intensity ranges by value. 
The proximity of single-cell data to the inspected area and the H\&E channel comparison, which is limited or non-existent in other tools, were described as particularly useful. The ability to simultaneously inspect and mark different regions of the tumor was perceived as a promising area for further development.

\subsection{Use Case 2: Lung Adenocarcinoma}

Lung adenocarcinoma is a common subset of lung cancer that, in later stages, often does not respond to chemotherapy. Immunotherapy has shown great promise, but patient response varies according to each tumor’s specific microenvironment. To assess why only certain patients respond, one needs an in-depth understanding of the tumor and immune landscape. Together with two biologists, we apply Scope2Screen to explore the immune reactions in lymphocyte structures.

\noindent\textbf{Data: }Using t-CyCIF, we have prepared a dataset of human lung adenocarcinoma wedge resection. The image data consists of 38 channels, each with a resolution of 39,843 x 29,227 pixels (12,949 x 9,499 microns) and 118.43 GB in size containing 534,677 segmented cells. The data is enriched with cell-type classification from a deep immune profiling of the tumor microenvironment.

\noindent\textbf{Analysis:}
We began the analysis with the whole tissue in the viewport and activated the DNA channel for a general overview of the tissue architecture. Addressing the feedback from Use Case 1, we equipped the lens with a set of four predefined channel combinations, designed to investigate the tumor's immune response from a high level to detail.

The first channel combination (Fig.~\ref{fig:multimodal_lung} A) consisted of basic cell markers and was designed to assess overall tissue composition. Lung cells are marked by TTF-1,
which enabled the detection of the aberrant tumor morphology in the upper region of the sample. The lens magnifier allowed the biologists to see that the tumor is heavily infiltrated with immune cells (marked by CD45 and Vimentin).  
We then switched to a second channel combination for immune cell markers (Fig.~\ref{fig:multimodal_lung} B) to further investigate the types of immune cells infiltrating the tumor. This revealed dense aggregates of immune cells, mainly composed of a core of B cells, surrounded by T cells, which are known tumor-associated tertiary lymphoid structures (TLS). Lens magnification allowed our experts to quickly detect, mark and further characterize the large number (10-12) of TLS's in this lung specimen. By switching to a third channel composition customized for TLS markers (Fig.~\ref{fig:multimodal_lung} C), we inspected each TLS more closely (see Sec.~\ref{sec:UseCase1} for a closer description).

We used the Dotter’s snapshot capability and placed landmarks as we reviewed the TLS's. We measured the TLS size using the ruler functionality (1,200 microns on average). We activated the cell type lens equipped with immune and cancer type classifications which showed B cells in the center (Fig.~\ref{fig:featureAugmentation} F, Fig.~\ref{fig:multimodal_lung} C) 
surrounded by epithelial lung cells. The cell biologists also used this setting to check the dataset segmentation, which they rated as very accurate. While focusing on one of these TLS, we activated HistoSearch to find areas with similar marker patterns, successfully identifying the other TLS areas. Surprisingly, in some TLS the HistoSearch detected the structure’s outer rim but not its center. Using the magnifier to zoom into these regions, we noted a distinctly higher level of B cell marker CD20. We focused the lens on this area and magnified for close-up inspection. To better understand the marker distribution in this region, we switched the lens function to the radial distribution chart (Fig.~\ref{fig:featureAugmentation} B), revealing a high Ki67 and PCNA concentration, markers for cell proliferation and growth.

Subsequently, we switched to the fourth channel group to compare lymphocyte markers across a TLS (Fig.~\ref{fig:multimodal_lung} D). We activated the histograms and moved the lens outside of a TLS towards the lung epithelial cells, recognizing TTF-1 positive tumor cells. Some of these were MHC-II positive. 
By shrinking the lens scope to single-cell size, we compared the TTF-1 cells, finding that cells with MHC-II expression are non-proliferative. This suggests that transient MHC-II expression is coupled with entry into a non-proliferative state.

\noindent\textbf{Feedback:}
While users were able to make meaningful observations throughout the evaluation, their commentary indicated two categories of improvements. 
Channel views and statistical views could be merged so users do not have to rely on rapid memory recall required for toggling filters. For example, simultaneous access to the single-cell marker intensity distribution would have been helpful to monitor non-selected markers for early exploration of broad immune cell lineages. 
Acknowledging that our tools do not support a wide range of unanticipated biological questions, the absence of tools for spatial analysis stalled certain leading lines of inquiry. Measuring the degree of cell proliferation around immune cell clusters would have been an important next step for our users, who recommended that we prioritize the introduction of supporting algorithms and visualizations for spatial correlation.

\vspace{-0.5em}
\section{Hands-On User Study and Questionnaire}
\label{sec:handson}
To further evaluate the intuitiveness and usefulness of Scope2Screen's interactive interfaces and lens features, we conducted hands-on user studies in which three of our four domain collaborators (that were involved in goal specification, iterative testing and design, and use-cases), two biologists and one pathologist, steered the tool, gave think-aloud feedback, and additionally filled out a questionnaire.

\noindent\textbf{Study Setup.} The sessions were conducted via Zoom with one subject matter expert (SME) at a time and two members from our visualization team. Scope2Screen was installed on a remote machine to which the experts had access. We used a ``think aloud'' approach~\cite{carpendale2008evaluating} as an opportunity for users to share feedback from their own interaction with the application. 
We recorded video and audio. The users worked sequentially through a list of Scope2Screen's features before freely exploring with the features practiced, following regions of biological interest. 
Sessions 
took between 70 and 90 minutes. 
They then filled out a questionnaire rating the usefulness of individual features with a 5 bin Likert scale (strongly agree to strongly disagree).

\noindent\textbf{Study and Feedback.} At first, the experts were asked to make use of global viewer features such as toggling image channels, panning \& zooming, and setting color ranges. Overall, they rated 
the application interface as intuitive and accessible (strongly agree, agree, and neutral). All agreed that focus+context 
improved exploration over a pure O+D approach and strongly agreed that the lens magnifier is helpful for observing local regions.  
Aligning with use-case feedback, the experts preferred the normal magnifier and found the plateau lens helpful to overcome distortion of the fish-eye but less intuitive. The pathologist preferred the circular shape for exploration and the rectangular lens for snapshots.
All experts agreed that the snapshot capabilities were extremely helpful but in some situations, the overlay can occlude areas underneath, hence functionality to show or hide them is needed. The dotting panel and annotation capabilities were used intermittently during evaluation and were rated helpful (2x strongly agree, 1x agree). The selection and combination of distinct channels (Section~\ref{sec:ChannelAnaysis}) within the lens achieved the same rating. One expert mentioned that these were especially beneficial for checking biases in assigned cell types, and two experts suggested to provide means to store channel combinations and color settings of the lens for repetitive analysis tasks. Most liked of these features was the split-screen lens to relate, e.g, CyCIF to H\&E data, and to validate alignment (3x strongly agree). The feature-augmentation lenses (Section~\ref{sec:featureAugmentation}) were also rated to be helpful (1x agree, 2x strongly agree). The linear histograms were preferred over the radial chart as they were easier to comprehend, but one expert mentioned that the radial chart provided a good relative overview of the distribution and might lead to unexpected discoveries. Another comment proposed that showing numbers, in addition to the visual encoding, would be helpful.  
Using the cell type lens, one expert said that the tooltip, similar as provided for the histogram and radial chart, should supplement the local cell type counts with cell type counts from the whole tissue in order to aid comparison and give context. It should further be possible to hide cell types. Lastly, users agreed that similarity search results (Section~\ref{sec:similarRegions}) align with a visual similarity impression, with slightly more conservative feedback (2x agree, 1x neutral), mostly due to the nature of the sensitivity threshold, which requires repetitive fine-tuning depending on the underlying image area. This could be improved with parameter optimization. One expert commented that ``what is considered similar'' depends on the morphological or molecular questions in focus, and hence the lens could be equipped with additional similarity methods. Overall, the experts found the tool ``easy to learn and use'' (2x agree, 1x strongly agree).

\vspace{-0.5em}
\section{Conclusion and Future Work}
We present a design study aimed at supporting single-cell research into the composition, molecular states, and phenotypes of normal and diseased tissues, a rapidly growing area of basic and translational biomedical research, as well as pathologists studying human tissues for the purpose of diagnosis and disease management.  Our Scope2Screen system supports fluid interactivity based on familiar microscopy metaphors while enabling multivariate exploration of quantitative and image data using a lens. 
Throughout the design process and expert feedback, we identified three key areas in which current work could be most usefully extended. 
\noindent\textbf{Combining Vision and Statistics: }
According to our experts, visual needs tools for presenting numerical data alongside image data. In many cases, generation of the numerical data is not the problem: computational biologists are familiar with scripting and statistical tools  for deriving single-cell data from images (via segmentation) and linking it to external sources (e.g., genomic data) but the information is most useful alongside the original images. Pathologists in particular need to combine deep knowledge of tissue architecture with quantitative data. However, most visual tools do not offer sufficient flexibility, and scripts or notebooks (e.g., Jupyter) lack the interactive visual exploration. We intend to extend Scope2Screen to support scripted statistical queries integrated with lensing.  
\noindent\textbf{High-dimensional Features on the Horizon: }
Recent development in digital imaging such as the ability to measure spatial distribution of RNA expression will result in data with thousands, not dozens, of  dimensions. Mapping such data into the image space while extracting relevant information will require dimensional reduction techniques and suitable visual representations of found features so that only the most relevant or explanatory data are presented. Very high-resolution 3D microscopy of tissues is also being integrated with the high-plex 2D data described here and this will require appropriate visual metaphors for moving between resolutions and data modalities.
\noindent\textbf{Scalability across Datasets: }
Our use cases demonstrate uses for Scope2Screen in the analysis of a single dataset stored locally.  However, digital histology is expected to transition to the analysis of multiple datasets accessed via the cloud. While Scope2Screen scales to a large set of images, it does not yet support analysis and annotation of image collections or work interactively with Docker-based image analysis pipelines.
Adding this functionality will close the gap from data exploration and analysis to generation of machine-assisted interpretative data reports for research and clinical applications including interactive publication via tools such as Minerva~\cite{rashid_interpretative_2020,hoffer_minerva_2020}.

\section{Acknowledgements}
This work is supported by the Ludwig
Center at Harvard Medical School, and by NIH/NCI grants NCI U2C-CA233262, NCI U2C-CA233280 and NCI U54-CA225088.

\section{Outside (Competing) Interests}
PKS is a member of the SAB or Board of Directors of Glencoe Software, Applied Biomath, and RareCyte Inc. and has equity in these companies; PKS is also a member of the SAB of NanoString.  Sorger declares that none of these relationships have influenced the content of this manuscript. Other authors declare no competing interests.

\bibliographystyle{abbrv-doi}

\bibliography{Lensing.bib}

\begin{thebibliography}{10}

\bibitem{jared_lensing}
Lensing, an npm package, https://www.npmjs.com/package/lensing, last accessed:
  8/06/2021.

\bibitem{noauthor_ome-tiff_nodate}
The {OME}-{TIFF} format — {OME} {Data} {Model} and {File} {Formats} 5.6.3
  documentation - https://docs.openmicroscopy.org/ome-model/5.6.3/ome-tiff/,
  last accessed: 3/31/2021.

\bibitem{noauthor_openseadragon_nodate}
{OpenSeadragon} - {An} open-source, web-based viewer for high-resolution
  zoomable images - https://openseadragon.github.io, last accessed: 3/31/2021.

\bibitem{scope2screen}
Scope2screen codebase, https://github.com/labsyspharm/scope2screen, last
  accessed: 8/06/2021.

\bibitem{noauthor_sliding_nodate}
Sliding window histogram, skimage: image processing in python, v0.18.0 docs -
  scikit-image.org/docs/stable/auto\_examples/features\_detection/plot\_
  windowed\_histogram.html, last accessed 3/31/2021.

\bibitem{tissue_align}
Visiopharm {TissueAlign} - high-quality alignment of serial sections -
  https://visiopharm.com/visiopharm-digital-image-analysis-software-features/tissuealign,
  last accessed: 6/30/2021.

\bibitem{visopharm}
Visiopharm {Viewer} - {AI}-driven {Pathology} software -
  https://visiopharm.com/visiopharm-digital-image-analysis-software-features/viewer,
  last accessed: 6/30/2021.

\bibitem{noauthor_webgl_nodate}
{WebGL} 2.0 {Specification}, https://www.khronos.org/registry/webgl/specs
  /latest/2.0/, last accessed: 3/31/2021.

\bibitem{noauthor_zarr_nodate}
Zarr — zarr 2.6.1 documentation, https://zarr.readthedocs.io/en/stable/, last
  accessed: 3/31/2021.

\bibitem{in_e_1984}
E.~H. Adelson, C.~H. Anderson, J.~R. Bergen, P.~J. Burt, and J.~M. Ogden.
\newblock Pyramid methods in image processing.
\newblock {\em RCA engineer}, 29(6):33--41, 1984.

\bibitem{allan_omero_2012}
C.~Allan, J.-M. Burel, J.~Moore, C.~Blackburn, M.~Linkert, S.~Loynton, et~al.
\newblock {OMERO}: flexible, model-driven data management for experimental
  biology.
\newblock {\em Nature Methods}, 9(3):245--253, Mar. 2012. doi: {{%
10\hspace{.1pt}\discretionary{.}{%
}{.}\hspace{.4pt}1038\discretionary{/}{%
}{/}nmeth\hspace{.1pt}\discretionary{.}{%
}{.}\hspace{.4pt}1896}}


\bibitem{arias-hernandez_pair_2011}
R.~Arias-Hernandez, L.~T. Kaastra, T.~M. Green, and B.~Fisher.
\newblock Pair {Analytics}: {Capturing} {Reasoning} {Processes} in
  {Collaborative} {Visual} {Analytics}.
\newblock In {\em 2011 44th {Hawaii} {International} {Conference} on {System}
  {Sciences}}, pp. 1--10, Jan. 2011.
\newblock ISSN: 1530-1605. doi: {{%
10\hspace{.1pt}\discretionary{.}{%
}{.}\hspace{.4pt}1109\discretionary{/}{%
}{/}HICSS\hspace{.1pt}\discretionary{.}{%
}{.}\hspace{.4pt}2011\hspace{.1pt}\discretionary{.}{%
}{.}\hspace{.4pt}339}}


\bibitem{bankhead_qupath_2017}
P.~Bankhead, M.~B. Loughrey, J.~A. Fernández, Y.~Dombrowski, D.~G. McArt,
  P.~D. Dunne, S.~McQuaid, et~al.
\newblock {QuPath}: {Open} source software for digital pathology image
  analysis.
\newblock {\em Scientific Reports}, 7(1):16878, Dec. 2017. doi: {{%
10\hspace{.1pt}\discretionary{.}{%
}{.}\hspace{.4pt}1038\discretionary{/}{%
}{/}s41598\discretionary{%
}{-}{-}017\discretionary{%
}{-}{-}17204\discretionary{%
}{-}{-}5}}


\bibitem{bier_toolglass_nodate}
E.~A. Bier, M.~C. Stone, K.~Pier, W.~Buxton, and T.~D. DeRose.
\newblock Toolglass and {Magic} {Lenses}: {The} {See}-{Through} {Interface}.
\newblock p.~8.

\bibitem{borovec_anhir_2020}
J.~Borovec, J.~Kybic, I.~Arganda-Carreras, D.~V. Sorokin, G.~Bueno, A.~V.
  Khvostikov, S.~Bakas, et~al.
\newblock {ANHIR}: {Automatic} {Non}-{Rigid} {Histological} {Image}
  {Registration} {Challenge}.
\newblock {\em IEEE Transactions on Medical Imaging}, 39(10):3042--3052, Oct.
  2020. doi: {{%
10\hspace{.1pt}\discretionary{.}{%
}{.}\hspace{.4pt}1109\discretionary{/}{%
}{/}TMI\hspace{.1pt}\discretionary{.}{%
}{.}\hspace{.4pt}2020\hspace{.1pt}\discretionary{.}{%
}{.}\hspace{.4pt}2986331}}


\bibitem{campanella_clinical-grade_2019}
G.~Campanella, M.~G. Hanna, L.~Geneslaw, A.~Miraflor, V.~Werneck Krauss~Silva,
  K.~J. Busam, et~al.
\newblock Clinical-grade computational pathology using weakly supervised deep
  learning on whole slide images.
\newblock {\em Nature Medicine}, 25(8):1301--1309, Aug. 2019.
\newblock Number: 8 Publisher: Nature Publishing Group. doi: {{%
10\hspace{.1pt}\discretionary{.}{%
}{.}\hspace{.4pt}1038\discretionary{/}{%
}{/}s41591\discretionary{%
}{-}{-}019\discretionary{%
}{-}{-}0508\discretionary{%
}{-}{-}1}}


\bibitem{carpendale_extending_1997}
M.~Carpendale, D.~Cowperthwaite, and F.~Fracchia.
\newblock Extending distortion viewing from {2D} to {3D}.
\newblock {\em IEEE Computer Graphics and Applications}, 17(4):42--51, Aug.
  1997. doi: {{%
10\hspace{.1pt}\discretionary{.}{%
}{.}\hspace{.4pt}1109\discretionary{/}{%
}{/}38\hspace{.1pt}\discretionary{.}{%
}{.}\hspace{.4pt}595268}}


\bibitem{carpendale2008evaluating}
S.~Carpendale.
\newblock Evaluating information visualizations.
\newblock In {\em Information visualization}, pp. 19--45. Springer, 2008.

\bibitem{chapman_scope_2020}
J.~A. Chapman, L.~M.~J. Lee, and N.~T. Swailes.
\newblock From {Scope} to {Screen}: {The} {Evolution} of {Histology}
  {Education}.
\newblock In P.~M. Rea, ed., {\em Biomedical {Visualisation}: {Volume} 8},
  Advances in {Experimental} {Medicine} and {Biology}, pp. 75--107. Springer
  International Publishing, Cham, 2020. doi: {{%
10\hspace{.1pt}\discretionary{.}{%
}{.}\hspace{.4pt}1007\discretionary{/}{%
}{/}978\discretionary{%
}{-}{-}3\discretionary{%
}{-}{-}030\discretionary{%
}{-}{-}47483\discretionary{%
}{-}{-}6\_5}}


\bibitem{cockburn_review_2009}
A.~Cockburn, A.~Karlson, and B.~B. Bederson.
\newblock A review of overview+detail, zooming, and focus+context interfaces.
\newblock {\em ACM Computing Surveys}, 41(1):2:1--2:31, Jan. 2009. doi: {{%
10\hspace{.1pt}\discretionary{.}{%
}{.}\hspace{.4pt}1145\discretionary{/}{%
}{/}1456650\hspace{.1pt}\discretionary{.}{%
}{.}\hspace{.4pt}1456652}}


\bibitem{dinkla_screenit_2017-1}
K.~Dinkla, H.~Strobelt, B.~Genest, S.~Reiling, M.~Borowsky, and H.~Pfister.
\newblock Screenit: {Visual} {Analysis} of {Cellular} {Screens}.
\newblock {\em IEEE Transactions on Visualization and Computer Graphics},
  23(1):591--600, Jan. 2017. doi: {{%
10\hspace{.1pt}\discretionary{.}{%
}{.}\hspace{.4pt}1109\discretionary{/}{%
}{/}TVCG\hspace{.1pt}\discretionary{.}{%
}{.}\hspace{.4pt}2016\hspace{.1pt}\discretionary{.}{%
}{.}\hspace{.4pt}2598587}}


\bibitem{dolatshah_ball-tree_2015}
M.~Dolatshah, A.~Hadian, and B.~Minaei-Bidgoli.
\newblock Ball*-tree: {Efficient} spatial indexing for constrained
  nearest-neighbor search in metric spaces.
\newblock {\em arXiv:1511.00628 [cs]}, Nov. 2015.
\newblock arXiv: 1511.00628.

\bibitem{gasteiger_flowlens_2011}
R.~Gasteiger, M.~Neugebauer, O.~Beuing, and B.~Preim.
\newblock The {FLOWLENS}: {A} {Focus}-and-{Context} {Visualization} {Approach}
  for {Exploration} of {Blood} {Flow} in {Cerebral} {Aneurysms}.
\newblock {\em IEEE Transactions on Visualization and Computer Graphics},
  17(12):2183--2192, Dec. 2011. doi: {{%
10\hspace{.1pt}\discretionary{.}{%
}{.}\hspace{.4pt}1109\discretionary{/}{%
}{/}TVCG\hspace{.1pt}\discretionary{.}{%
}{.}\hspace{.4pt}2011\hspace{.1pt}\discretionary{.}{%
}{.}\hspace{.4pt}243}}


\bibitem{gehlenborg_vitessce_2021}
N.~Gehlenborg, T.~Manz, and I.~Gold.
\newblock Vitessce - {A} {Framework} and {Visual} {Integration} {Tool} for
  {Exploration} of (spatial) single-cell experiment data., 2021.
\newblock OSF Preprints. August 12. doi:10.31219/osf.io/wd2gu.

\bibitem{gerdes_highly_2013}
M.~J. Gerdes, C.~J. Sevinsky, A.~Sood, S.~Adak, M.~O. Bello, A.~Bordwell,
  et~al.
\newblock Highly multiplexed single-cell analysis of formalin-fixed,
  paraffin-embedded cancer tissue.
\newblock {\em Proceedings of the National Academy of Sciences},
  110(29):11982--11987, July 2013.
\newblock Publisher: National Academy of Sciences Section: Biological Sciences.
  doi: {{%
10\hspace{.1pt}\discretionary{.}{%
}{.}\hspace{.4pt}1073\discretionary{/}{%
}{/}pnas\hspace{.1pt}\discretionary{.}{%
}{.}\hspace{.4pt}1300136110}}


\bibitem{goltsev_deep_2018}
Y.~Goltsev, N.~Samusik, J.~Kennedy-Darling, S.~Bhate, M.~Hale, G.~Vazquez,
  S.~Black, and G.~P. Nolan.
\newblock Deep {Profiling} of {Mouse} {Splenic} {Architecture} with {CODEX}
  {Multiplexed} {Imaging}.
\newblock {\em Cell}, 174(4):968--981.e15, Aug. 2018. doi: {{%
10\hspace{.1pt}\discretionary{.}{%
}{.}\hspace{.4pt}1016\discretionary{/}{%
}{/}j\hspace{.1pt}\discretionary{.}{%
}{.}\hspace{.4pt}cell\hspace{.1pt}\discretionary{.}{%
}{.}\hspace{.4pt}2018\hspace{.1pt}\discretionary{.}{%
}{.}\hspace{.4pt}07\hspace{.1pt}\discretionary{.}{%
}{.}\hspace{.4pt}010}}


\bibitem{gutwin_fisheyes_2003}
C.~Gutwin and A.~Skopik.
\newblock Fisheyes are good for large steering tasks.
\newblock In {\em Proceedings of the {SIGCHI} {Conference} on {Human} {Factors}
  in {Computing} {Systems}}, {CHI} '03, pp. 201--208. Association for Computing
  Machinery, New York, NY, USA, Apr. 2003. doi: {{%
10\hspace{.1pt}\discretionary{.}{%
}{.}\hspace{.4pt}1145\discretionary{/}{%
}{/}642611\hspace{.1pt}\discretionary{.}{%
}{.}\hspace{.4pt}642648}}


\bibitem{hoffer_minerva_2020}
J.~Hoffer, R.~Rashid, J.~L. Muhlich, Y.-A. Chen, D.~P.~W. Russell, J.~Ruokonen,
  R.~Krueger, H.~Pfister, S.~Santagata, and P.~K. Sorger.
\newblock Minerva: a light-weight, narrative image browser for multiplexed
  tissue images.
\newblock {\em Journal of Open Source Software}, 5(54):2579, Oct. 2020. doi:
  {{%
10\hspace{.1pt}\discretionary{.}{%
}{.}\hspace{.4pt}21105\discretionary{/}{%
}{/}joss\hspace{.1pt}\discretionary{.}{%
}{.}\hspace{.4pt}02579}}


\bibitem{hurter_moleview_2011}
C.~Hurter, A.~Telea, and O.~Ersoy.
\newblock {MoleView}: {An} {Attribute} and {Structure}-{Based} {Semantic}
  {Lens} for {Large} {Element}-{Based} {Plots}.
\newblock {\em IEEE Transactions on Visualization and Computer Graphics},
  17(12):2600--2609, Dec. 2011. doi: {{%
10\hspace{.1pt}\discretionary{.}{%
}{.}\hspace{.4pt}1109\discretionary{/}{%
}{/}TVCG\hspace{.1pt}\discretionary{.}{%
}{.}\hspace{.4pt}2011\hspace{.1pt}\discretionary{.}{%
}{.}\hspace{.4pt}223}}


\bibitem{indica_labs_halo_nodate}
{Indica Labs}.
\newblock {HALO} - image analysis platform for quantitative tissue analysis in
  digital pathology - https://indicalab.com/halo/, last-accessed: 7/1/2021.

\bibitem{ip_saliency-assisted_2011}
C.~Y. Ip and A.~Varshney.
\newblock Saliency-{Assisted} {Navigation} of {Very} {Large} {Landscape}
  {Images}.
\newblock {\em IEEE Transactions on Visualization and Computer Graphics},
  17(12):1737--1746, Dec. 2011. doi: {{%
10\hspace{.1pt}\discretionary{.}{%
}{.}\hspace{.4pt}1109\discretionary{/}{%
}{/}TVCG\hspace{.1pt}\discretionary{.}{%
}{.}\hspace{.4pt}2011\hspace{.1pt}\discretionary{.}{%
}{.}\hspace{.4pt}231}}


\bibitem{klein_elastix_2010}
S.~Klein, M.~Staring, K.~Murphy, M.~A. Viergever, and J.~P.~W. Pluim.
\newblock elastix: {A} {Toolbox} for {Intensity}-{Based} {Medical} {Image}
  {Registration}.
\newblock {\em IEEE Transactions on Medical Imaging}, 29(1):196--205, Jan.
  2010. doi: {{%
10\hspace{.1pt}\discretionary{.}{%
}{.}\hspace{.4pt}1109\discretionary{/}{%
}{/}TMI\hspace{.1pt}\discretionary{.}{%
}{.}\hspace{.4pt}2009\hspace{.1pt}\discretionary{.}{%
}{.}\hspace{.4pt}2035616}}


\bibitem{krueger_facetto_2020}
R.~Krueger, J.~Beyer, W.-D. Jang, N.~W. Kim, A.~Sokolov, P.~K. Sorger, and
  H.~Pfister.
\newblock Facetto: {Combining} {Unsupervised} and {Supervised} {Learning} for
  {Hierarchical} {Phenotype} {Analysis} in {Multi}-{Channel} {Image} {Data}.
\newblock {\em IEEE Transactions on Visualization and Computer Graphics},
  26(1):227--237, Jan. 2020. doi: {{%
10\hspace{.1pt}\discretionary{.}{%
}{.}\hspace{.4pt}1109\discretionary{/}{%
}{/}TVCG\hspace{.1pt}\discretionary{.}{%
}{.}\hspace{.4pt}2019\hspace{.1pt}\discretionary{.}{%
}{.}\hspace{.4pt}2934547}}


\bibitem{lekschas_pattern-driven_2020}
F.~Lekschas, M.~Behrisch, B.~Bach, P.~Kerpedjiev, N.~Gehlenborg, and
  H.~Pfister.
\newblock Pattern-{Driven} {Navigation} in {2D} {Multiscale} {Visualizations}
  with {Scalable} {Insets}.
\newblock {\em IEEE Transactions on Visualization and Computer Graphics},
  26(1):611--621, Jan. 2020. doi: {{%
10\hspace{.1pt}\discretionary{.}{%
}{.}\hspace{.4pt}1109\discretionary{/}{%
}{/}TVCG\hspace{.1pt}\discretionary{.}{%
}{.}\hspace{.4pt}2019\hspace{.1pt}\discretionary{.}{%
}{.}\hspace{.4pt}2934555}}


\bibitem{lin_cyclic_2016}
J.-R. Lin, M.~Fallahi‐Sichani, J.-Y. Chen, and P.~K. Sorger.
\newblock Cyclic {Immunofluorescence} ({CycIF}), {A} {Highly} {Multiplexed}
  {Method} for {Single}-cell {Imaging}.
\newblock {\em Current Protocols in Chemical Biology}, 8(4):251--264, 2016.
\newblock \_eprint:
  https://currentprotocols.onlinelibrary.wiley.com/doi/pdf/10.1002/cpch.14.
  doi: {{%
10\hspace{.1pt}\discretionary{.}{%
}{.}\hspace{.4pt}1002\discretionary{/}{%
}{/}cpch\hspace{.1pt}\discretionary{.}{%
}{.}\hspace{.4pt}14}}


\bibitem{lin_highly_nodate}
J.-R. Lin, B.~Izar, S.~Wang, C.~Yapp, S.~Mei, P.~M. Shah, S.~Santagata, and
  P.~K. Sorger.
\newblock Highly multiplexed immunofluorescence imaging of human tissues and
  tumors using t-{CyCIF} and conventional optical microscopes.
\newblock {\em eLife}, 7. doi: {{%
10\hspace{.1pt}\discretionary{.}{%
}{.}\hspace{.4pt}7554\discretionary{/}{%
}{/}eLife\hspace{.1pt}\discretionary{.}{%
}{.}\hspace{.4pt}31657}}


\bibitem{liu_comparison_2019}
X.~Liu, W.~Song, B.~Y. Wong, T.~Zhang, S.~Yu, G.~N. Lin, and X.~Ding.
\newblock A comparison framework and guideline of clustering methods for mass
  cytometry data.
\newblock {\em Genome Biology}, 20(1):297, Dec. 2019. doi: {{%
10\hspace{.1pt}\discretionary{.}{%
}{.}\hspace{.4pt}1186\discretionary{/}{%
}{/}s13059\discretionary{%
}{-}{-}019\discretionary{%
}{-}{-}1917\discretionary{%
}{-}{-}7}}


\bibitem{lsp_labsyspharms3segmenter_221}
{LSP}.
\newblock labsyspharm/s3segmenter - segmentation of cells from probability maps
  - https://hub.docker.com/r/labsyspharm/s3segmenter, last accessed: 6/30/2021,
  Mar. 221.
\newblock https://hub.docker.com/r/labsyspharm/s3segmenter.

\bibitem{mackinlay_automating_1986}
J.~Mackinlay.
\newblock Automating the design of graphical presentations of relational
  information.
\newblock {\em ACM Transactions on Graphics}, 5(2):110--141, Apr. 1986. doi:
  {{%
10\hspace{.1pt}\discretionary{.}{%
}{.}\hspace{.4pt}1145\discretionary{/}{%
}{/}22949\hspace{.1pt}\discretionary{.}{%
}{.}\hspace{.4pt}22950}}


\bibitem{madabhushi_image_2016}
A.~Madabhushi and G.~Lee.
\newblock Image analysis and machine learning in digital pathology:
  {Challenges} and opportunities.
\newblock {\em Medical Image Analysis}, 33:170--175, Oct. 2016. doi: {{%
10\hspace{.1pt}\discretionary{.}{%
}{.}\hspace{.4pt}1016\discretionary{/}{%
}{/}j\hspace{.1pt}\discretionary{.}{%
}{.}\hspace{.4pt}media\hspace{.1pt}\discretionary{.}{%
}{.}\hspace{.4pt}2016\hspace{.1pt}\discretionary{.}{%
}{.}\hspace{.4pt}06\hspace{.1pt}\discretionary{.}{%
}{.}\hspace{.4pt}037}}


\bibitem{manz_viv_2020}
T.~Manz, I.~Gold, N.~H. Patterson, C.~McCallum, M.~S. Keller, B.~W.~H. Ii,
  K.~Börner, J.~M. Spraggins, and N.~Gehlenborg.
\newblock Viv: {Multiscale} {Visualization} of {High}-{Resolution}
  {Multiplexed} {Bioimaging} {Data} on the {Web}.
\newblock Technical report, OSF Preprints, Aug. 2020.
\newblock type: article. doi: {{%
10\hspace{.1pt}\discretionary{.}{%
}{.}\hspace{.4pt}31219\discretionary{/}{%
}{/}osf\hspace{.1pt}\discretionary{.}{%
}{.}\hspace{.4pt}io\discretionary{/}{%
}{/}wd2gu}}


\bibitem{marstal_simpleelastix_2016}
K.~Marstal, F.~Berendsen, M.~Staring, and S.~Klein.
\newblock {SimpleElastix}: {A} {User}-{Friendly}, {Multi}-lingual {Library} for
  {Medical} {Image} {Registration}.
\newblock In {\em 2016 {IEEE} {Conference} on {Computer} {Vision} and {Pattern}
  {Recognition} {Workshops} ({CVPRW})}, pp. 574--582, June 2016. doi: {{%
10\hspace{.1pt}\discretionary{.}{%
}{.}\hspace{.4pt}1109\discretionary{/}{%
}{/}CVPRW\hspace{.1pt}\discretionary{.}{%
}{.}\hspace{.4pt}2016\hspace{.1pt}\discretionary{.}{%
}{.}\hspace{.4pt}78}}


\bibitem{miao_quick_nodate}
R.~Miao, R.~Toth, Y.~Zhou, A.~Madabhushi, and A.~Janowczyk.
\newblock Quick {Annotator}: an open-source digital pathology based rapid image
  annotation tool.
\newblock p.~14.

\bibitem{microsoft_silverlight_2021}
{Microsoft}.
\newblock ({Silverlight}) {Deep} {Zoom} {File} {Format} {Overview},
  https://docs.microsoft.com/en-us/previous-versions/windows/
  silverlight/dotnet-windows-silverlight/cc645077(v=vs.95), last accessed:
  3/31/2021, Oct. 2021.

\bibitem{alistair_miles_zarr-developerszarr-python_2020}
A.~Miles, J.~Kirkham, M.~Durant, J.~Bourbeau, T.~Onalan, J.~Hamman, et~al.
\newblock zarr-developers/zarr-python: v2.4.0, Jan. 2020. doi: {{%
10\hspace{.1pt}\discretionary{.}{%
}{.}\hspace{.4pt}5281\discretionary{/}{%
}{/}zenodo\hspace{.1pt}\discretionary{.}{%
}{.}\hspace{.4pt}3773450}}


\bibitem{mindek_managing_2014}
P.~Mindek, M.~E. Gröller, and S.~Bruckner.
\newblock Managing {Spatial} {Selections} {With} {Contextual} {Snapshots}.
\newblock {\em Computer Graphics Forum}, 33(8):132--144, 2014.
\newblock \_eprint: https://onlinelibrary.wiley.com/doi/pdf/10.1111/cgf.12406.
  doi: {{%
10\hspace{.1pt}\discretionary{.}{%
}{.}\hspace{.4pt}1111\discretionary{/}{%
}{/}cgf\hspace{.1pt}\discretionary{.}{%
}{.}\hspace{.4pt}12406}}


\bibitem{molin_diagnostic_2016}
J.~Molin.
\newblock {\em Diagnostic {Review} with {Digital} {Pathology}}.
\newblock {PhD} {Thesis}, Chalmers Tekniska Hogskola (Sweden), 2016.

\bibitem{muhlich2021stitching}
J.~L. Muhlich, Y.-A. Chen, D.~P.~W. Russell, and P.~K. Sorger.
\newblock Stitching and registering highly multiplexed whole slide images of
  tissues and tumors using ashlar software.
\newblock {\em bioRxiv}, 2021.

\bibitem{muhlich_jeremy_et_al_labsyspharmashlar_2021}
{Muhlich, Jeremy et al.}
\newblock Ashlar - alignment by simultaneous harmonization of layer/adjacency
  registration - https://github.com/labsyspharm/ashlar, last accessed:
  6/30/2021, Mar. 2021.

\bibitem{morth_paraglyder_2020}
E.~Mörth, I.~S. Haldorsen, S.~Bruckner, and N.~N. Smit.
\newblock {ParaGlyder}: {Probe}-driven {Interactive} {Visual} {Analysis} for
  {Multiparametric} {Medical} {Imaging} {Data}.
\newblock In N.~Magnenat-Thalmann et~al., eds., {\em Advances in {Computer}
  {Graphics}}, Lecture {Notes} in {Computer} {Science}, pp. 351--363. Springer
  International Publishing, Cham, 2020. doi: {{%
10\hspace{.1pt}\discretionary{.}{%
}{.}\hspace{.4pt}1007\discretionary{/}{%
}{/}978\discretionary{%
}{-}{-}3\discretionary{%
}{-}{-}030\discretionary{%
}{-}{-}61864\discretionary{%
}{-}{-}3\_29}}


\bibitem{nirmal_et_al_scimap-_nodate}
{Nirmal et al.}
\newblock Scimap- {Single}-{Cell} {Image} {Analysis} {Package},
  https://scimap-doc.readthedocs.io/en/latest/, last accessed: 2021-08-08.

\bibitem{pallua_future_2020}
J.~D. Pallua, A.~Brunner, B.~Zelger, M.~Schirmer, and J.~Haybaeck.
\newblock The future of pathology is digital.
\newblock {\em Pathology - Research and Practice}, 216(9):153040, Sept. 2020.
  doi: {{%
10\hspace{.1pt}\discretionary{.}{%
}{.}\hspace{.4pt}1016\discretionary{/}{%
}{/}j\hspace{.1pt}\discretionary{.}{%
}{.}\hspace{.4pt}prp\hspace{.1pt}\discretionary{.}{%
}{.}\hspace{.4pt}2020\hspace{.1pt}\discretionary{.}{%
}{.}\hspace{.4pt}153040}}


\bibitem{perreault_median_2007}
S.~Perreault and P.~Hebert.
\newblock Median {Filtering} in {Constant} {Time}.
\newblock {\em IEEE Transactions on Image Processing}, 16(9):2389--2394, Sept.
  2007. doi: {{%
10\hspace{.1pt}\discretionary{.}{%
}{.}\hspace{.4pt}1109\discretionary{/}{%
}{/}TIP\hspace{.1pt}\discretionary{.}{%
}{.}\hspace{.4pt}2007\hspace{.1pt}\discretionary{.}{%
}{.}\hspace{.4pt}902329}}


\bibitem{porikli_integral_2005}
F.~Porikli.
\newblock Integral histogram: a fast way to extract histograms in {Cartesian}
  spaces.
\newblock In {\em 2005 {IEEE} {Computer} {Society} {Conference} on {Computer}
  {Vision} and {Pattern} {Recognition} ({CVPR}'05)}, vol.~1, pp. 829--836 vol.
  1, June 2005.
\newblock ISSN: 1063-6919. doi: {{%
10\hspace{.1pt}\discretionary{.}{%
}{.}\hspace{.4pt}1109\discretionary{/}{%
}{/}CVPR\hspace{.1pt}\discretionary{.}{%
}{.}\hspace{.4pt}2005\hspace{.1pt}\discretionary{.}{%
}{.}\hspace{.4pt}188}}


\bibitem{rashid_interpretative_2020}
R.~Rashid, Y.-A. Chen, J.~Hoffer, J.~L. Muhlich, J.-R. Lin, R.~Krueger,
  H.~Pfister, R.~Mitchell, S.~Santagata, and P.~K. Sorger.
\newblock Interpretative guides for interacting with tissue atlas and digital
  pathology data using the {Minerva} browser.
\newblock {\em bioRxiv}, p. 2020.03.27.001834, Mar. 2020.
\newblock Publisher: Cold Spring Harbor Laboratory Section: New Results. doi:
  {{%
10\hspace{.1pt}\discretionary{.}{%
}{.}\hspace{.4pt}1101\discretionary{/}{%
}{/}2020\hspace{.1pt}\discretionary{.}{%
}{.}\hspace{.4pt}03\hspace{.1pt}\discretionary{.}{%
}{.}\hspace{.4pt}27\hspace{.1pt}\discretionary{.}{%
}{.}\hspace{.4pt}001834}}


\bibitem{rozenblatt-rosen_human_2020}
O.~Rozenblatt-Rosen, A.~Regev, P.~Oberdoerffer, T.~Nawy, A.~Hupalowska, J.~E.
  Rood, O.~Ashenberg, E.~Cerami, R.~J. Coffey, E.~Demir, et~al.
\newblock The {Human} {Tumor} {Atlas} {Network}: {Charting} {Tumor}
  {Transitions} across {Space} and {Time} at {Single}-{Cell} {Resolution}.
\newblock {\em Cell}, 181(2):236--249, Apr. 2020. doi: {{%
10\hspace{.1pt}\discretionary{.}{%
}{.}\hspace{.4pt}1016\discretionary{/}{%
}{/}j\hspace{.1pt}\discretionary{.}{%
}{.}\hspace{.4pt}cell\hspace{.1pt}\discretionary{.}{%
}{.}\hspace{.4pt}2020\hspace{.1pt}\discretionary{.}{%
}{.}\hspace{.4pt}03\hspace{.1pt}\discretionary{.}{%
}{.}\hspace{.4pt}053}}


\bibitem{saunders_pancreatlas_2020}
D.~C. Saunders, J.~Messmer, I.~Kusmartseva, M.~L. Beery, M.~Yang, M.~A.
  Atkinson, A.~C. Powers, J.-P. Cartailler, and M.~Brissova.
\newblock Pancreatlas: {Applying} an {Adaptable} {Framework} to {Map} the
  {Human} {Pancreas} in {Health} and {Disease}.
\newblock {\em Patterns}, 1(8):100120, Nov. 2020. doi: {{%
10\hspace{.1pt}\discretionary{.}{%
}{.}\hspace{.4pt}1016\discretionary{/}{%
}{/}j\hspace{.1pt}\discretionary{.}{%
}{.}\hspace{.4pt}patter\hspace{.1pt}\discretionary{.}{%
}{.}\hspace{.4pt}2020\hspace{.1pt}\discretionary{.}{%
}{.}\hspace{.4pt}100120}}


\bibitem{schapiro_histocat_2017}
D.~Schapiro, H.~W. Jackson, S.~Raghuraman, and J.~R. Fischer.
\newblock {histoCAT}: analysis of cell phenotypes and interactions in multiplex
  image cytometry data.
\newblock {\em Nature methods}, 14(9):873--876, 2017.

\bibitem{schapiro_mcmicro_2021}
D.~Schapiro, A.~Sokolov, C.~Yapp, J.~L. Muhlich, J.~Hess, J.-R. Lin, Y.-A.
  Chen, M.~K. Nariya, G.~J. Baker, J.~Ruokonen, et~al.
\newblock {MCMICRO}: {A} scalable, modular image-processing pipeline for
  multiplexed tissue imaging.
\newblock {\em bioRxiv}, p. 2021.03.15.435473, Mar. 2021.
\newblock Publisher: Cold Spring Harbor Laboratory Section: New Results. doi:
  {{%
10\hspace{.1pt}\discretionary{.}{%
}{.}\hspace{.4pt}1101\discretionary{/}{%
}{/}2021\hspace{.1pt}\discretionary{.}{%
}{.}\hspace{.4pt}03\hspace{.1pt}\discretionary{.}{%
}{.}\hspace{.4pt}15\hspace{.1pt}\discretionary{.}{%
}{.}\hspace{.4pt}435473}}


\bibitem{sedlmair_design_2012}
M.~Sedlmair, M.~Meyer, and T.~Munzner.
\newblock Design {Study} {Methodology}: {Reflections} from the {Trenches} and
  the {Stacks}.
\newblock {\em IEEE Transactions on Visualization and Computer Graphics},
  18(12):2431--2440, Dec. 2012. doi: {{%
10\hspace{.1pt}\discretionary{.}{%
}{.}\hspace{.4pt}1109\discretionary{/}{%
}{/}TVCG\hspace{.1pt}\discretionary{.}{%
}{.}\hspace{.4pt}2012\hspace{.1pt}\discretionary{.}{%
}{.}\hspace{.4pt}213}}


\bibitem{shoemaker_supporting_2007}
G.~Shoemaker and C.~Gutwin.
\newblock Supporting multi-point interaction in visual workspaces.
\newblock In {\em Proceedings of the {SIGCHI} {Conference} on {Human} {Factors}
  in {Computing} {Systems}}, {CHI} '07, pp. 999--1008. Association for
  Computing Machinery, New York, NY, USA, Apr. 2007. doi: {{%
10\hspace{.1pt}\discretionary{.}{%
}{.}\hspace{.4pt}1145\discretionary{/}{%
}{/}1240624\hspace{.1pt}\discretionary{.}{%
}{.}\hspace{.4pt}1240777}}


\bibitem{nicholas_sofroniew_naparinapari_2021}
N.~Sofroniew, T.~Lambert, K.~Evans, P.~Winston, J.~Nunez-Iglesias, G.~Bokota,
  K.~Yamauchi, A.~C. Solak, ziyangczi, G.~Buckley, et~al.
\newblock napari/napari: 0.4.4rc0, Jan. 2021. doi: {{%
10\hspace{.1pt}\discretionary{.}{%
}{.}\hspace{.4pt}5281\discretionary{/}{%
}{/}zenodo\hspace{.1pt}\discretionary{.}{%
}{.}\hspace{.4pt}4470554}}


\bibitem{somarakis_visual_2021}
A.~Somarakis, M.~E. Ijsselsteijn, S.~J. Luk, B.~Kenkhuis, N.~F. C.~C.
  de~Miranda, B.~P.~F. Lelieveldt, and T.~Höllt.
\newblock Visual cohort comparison for spatial single-cell omics-data.
\newblock {\em IEEE Transactions on Visualization and Computer Graphics},
  27(2):733--743, Feb. 2021.
\newblock arXiv: 2006.05175. doi: {{%
10\hspace{.1pt}\discretionary{.}{%
}{.}\hspace{.4pt}1109\discretionary{/}{%
}{/}TVCG\hspace{.1pt}\discretionary{.}{%
}{.}\hspace{.4pt}2020\hspace{.1pt}\discretionary{.}{%
}{.}\hspace{.4pt}3030336}}


\bibitem{sommer_ilastik_2011}
C.~Sommer, C.~Straehle, U.~Köthe, and F.~A. Hamprecht.
\newblock Ilastik: {Interactive} learning and segmentation toolkit.
\newblock In {\em 2011 {IEEE} {International} {Symposium} on {Biomedical}
  {Imaging}: {From} {Nano} to {Macro}}, pp. 230--233, Mar. 2011.
\newblock ISSN: 1945-8452. doi: {{%
10\hspace{.1pt}\discretionary{.}{%
}{.}\hspace{.4pt}1109\discretionary{/}{%
}{/}ISBI\hspace{.1pt}\discretionary{.}{%
}{.}\hspace{.4pt}2011\hspace{.1pt}\discretionary{.}{%
}{.}\hspace{.4pt}5872394}}


\bibitem{stoltzfus_cytomap_2020}
C.~R. Stoltzfus, J.~Filipek, B.~H. Gern, B.~E. Olin, J.~M. Leal, Y.~Wu, et~al.
\newblock {CytoMAP}: {A} {Spatial} {Analysis} {Toolbox} {Reveals} {Features} of
  {Myeloid} {Cell} {Organization} in {Lymphoid} {Tissues}.
\newblock {\em Cell Reports}, 31(3):107523, Apr. 2020. doi: {{%
10\hspace{.1pt}\discretionary{.}{%
}{.}\hspace{.4pt}1016\discretionary{/}{%
}{/}j\hspace{.1pt}\discretionary{.}{%
}{.}\hspace{.4pt}celrep\hspace{.1pt}\discretionary{.}{%
}{.}\hspace{.4pt}2020\hspace{.1pt}\discretionary{.}{%
}{.}\hspace{.4pt}107523}}


\bibitem{stritt_orbit_2020}
M.~Stritt, A.~K. Stalder, and E.~Vezzali.
\newblock Orbit {Image} {Analysis}: {An} open-source whole slide image analysis
  tool.
\newblock {\em PLOS Computational Biology}, 16(2):e1007313, Feb. 2020.
\newblock Publisher: Public Library of Science. doi: {{%
10\hspace{.1pt}\discretionary{.}{%
}{.}\hspace{.4pt}1371\discretionary{/}{%
}{/}journal\hspace{.1pt}\discretionary{.}{%
}{.}\hspace{.4pt}pcbi\hspace{.1pt}\discretionary{.}{%
}{.}\hspace{.4pt}1007313}}


\bibitem{titford_long_2005}
M.~Titford.
\newblock The long history of hematoxylin.
\newblock {\em Biotechnic \& Histochemistry}, 80(2):73--78, Jan. 2005.
\newblock Publisher: Taylor \& Francis \_eprint:
  https://doi.org/10.1080/10520290500138372. doi: {{%
10\hspace{.1pt}\discretionary{.}{%
}{.}\hspace{.4pt}1080\discretionary{/}{%
}{/}10520290500138372}}


\bibitem{tominski_survey_2014}
C.~Tominski, S.~Gladisch, U.~Kister, R.~Dachselt, and H.~Schumann.
\newblock A {Survey} on {Interactive} {Lenses} in {Visualization}.
\newblock {\em EuroVis - STARs}, p. 20 pages, 2014.
\newblock Publisher: The Eurographics Association. doi: {{%
10\hspace{.1pt}\discretionary{.}{%
}{.}\hspace{.4pt}2312\discretionary{/}{%
}{/}EUROVISSTAR\hspace{.1pt}\discretionary{.}{%
}{.}\hspace{.4pt}20141172}}


\bibitem{tominski_interactive_2017}
C.~Tominski, S.~Gladisch, U.~Kister, R.~Dachselt, and H.~Schumann.
\newblock Interactive {Lenses} for {Visualization}: {An} {Extended} {Survey}.
\newblock {\em Computer Graphics Forum}, 36(6):173--200, 2017.
\newblock \_eprint: https://onlinelibrary.wiley.com/doi/pdf/10.1111/cgf.12871.
  doi: {{%
10\hspace{.1pt}\discretionary{.}{%
}{.}\hspace{.4pt}1111\discretionary{/}{%
}{/}cgf\hspace{.1pt}\discretionary{.}{%
}{.}\hspace{.4pt}12871}}


\bibitem{trapp_3d_2008}
M.~Trapp, T.~Glander, H.~Buchholz, and J.~Döllner.
\newblock {3D} {Generalization} {Lenses} for {Interactive} {Focus} + {Context}
  {Visualization} of {Virtual} {City} {Models}.
\newblock In {\em 2008 12th {International} {Conference} {Information}
  {Visualisation}}, pp. 356--361, July 2008.
\newblock ISSN: 2375-0138. doi: {{%
10\hspace{.1pt}\discretionary{.}{%
}{.}\hspace{.4pt}1109\discretionary{/}{%
}{/}IV\hspace{.1pt}\discretionary{.}{%
}{.}\hspace{.4pt}2008\hspace{.1pt}\discretionary{.}{%
}{.}\hspace{.4pt}18}}


\bibitem{van_wijk_smooth_2003}
J.~van Wijk and W.~Nuij.
\newblock Smooth and efficient zooming and panning.
\newblock In {\em {IEEE} {Symposium} on {Information} {Visualization} 2003
  ({IEEE} {Cat}. {No}.{03TH8714})}, pp. 15--23. IEEE, Seattle, WA, USA, 2003.
  doi: {{%
10\hspace{.1pt}\discretionary{.}{%
}{.}\hspace{.4pt}1109\discretionary{/}{%
}{/}INFVIS\hspace{.1pt}\discretionary{.}{%
}{.}\hspace{.4pt}2003\hspace{.1pt}\discretionary{.}{%
}{.}\hspace{.4pt}1249004}}


\bibitem{vollmer_hierarchical_2018}
J.~O. Vollmer, M.~Trapp, H.~Schumann, and J.~Döllner.
\newblock Hierarchical {Spatial} {Aggregation} for {Level}-of-{Detail}
  {Visualization} of {3D} {Thematic} {Data}.
\newblock {\em ACM Transactions on Spatial Algorithms and Systems}, 4(3):1--23,
  Sept. 2018. doi: {{%
10\hspace{.1pt}\discretionary{.}{%
}{.}\hspace{.4pt}1145\discretionary{/}{%
}{/}3234506}}


\bibitem{weinstein_cancer_2013}
J.~N. Weinstein, E.~A. Collisson, G.~B. Mills, K.~R.~M. Shaw, B.~A. Ozenberger,
  K.~Ellrott, I.~Shmulevich, C.~Sander, and J.~M. Stuart.
\newblock The {Cancer} {Genome} {Atlas} {Pan}-{Cancer} analysis project.
\newblock {\em Nature Genetics}, 45(10):1113--1120, Oct. 2013.
\newblock Number: 10 Publisher: Nature Publishing Group. doi: {{%
10\hspace{.1pt}\discretionary{.}{%
}{.}\hspace{.4pt}1038\discretionary{/}{%
}{/}ng\hspace{.1pt}\discretionary{.}{%
}{.}\hspace{.4pt}2764}}


\bibitem{weinstein_prospects_1986}
R.~S. Weinstein.
\newblock Prospects for telepathology.
\newblock {\em Human Pathology}, 17(5):433--434, May 1986. doi: {{%
10\hspace{.1pt}\discretionary{.}{%
}{.}\hspace{.4pt}1016\discretionary{/}{%
}{/}s0046\discretionary{%
}{-}{-}8177\discretionary{%
}{(}{(}86\discretionary{)}{%
}{)}80028\discretionary{%
}{-}{-}4}}


\bibitem{zhao_trailmap_2013}
J.~Zhao, D.~Wigdor, and R.~Balakrishnan.
\newblock {TrailMap}: facilitating information seeking in a multi-scale digital
  map via implicit bookmarking.
\newblock In {\em Proceedings of the {SIGCHI} {Conference} on {Human} {Factors}
  in {Computing} {Systems}}, pp. 3009--3018. ACM, Paris France, Apr. 2013. doi:
  {{%
10\hspace{.1pt}\discretionary{.}{%
}{.}\hspace{.4pt}1145\discretionary{/}{%
}{/}2470654\hspace{.1pt}\discretionary{.}{%
}{.}\hspace{.4pt}2481417}}


\end{thebibliography}
\end{document}